\documentclass[aps,prd,amsmath,amsfonts,nofootinbib,preprintnumbers, superscriptaddress,eqsecnum]{revtex4-2}
\pdfoutput=1
\usepackage{color,graphicx,slashed,multirow,amssymb,amsmath,multirow}
\usepackage{braket}
\usepackage{graphicx}
\usepackage[utf8]{inputenc}
\usepackage[section]{placeins}
\usepackage{xcolor,colortbl}
\usepackage{caption}
\usepackage{subcaption}
\usepackage{hyperref}
\usepackage[capitalise,noabbrev,sort&compress]{cleveref}

\usepackage[normalem]{ulem}

\usepackage[shortlabels]{enumitem}
\AtBeginDocument{\usepackage{booktabs}}
\renewcommand{\arraystretch}{1.2}

\def \met  {\mbox{${E\!\!\!\!/_T}$}}

\newcommand{\GeV}{{\rm\ GeV}}
\newcommand{\TeV}{{\rm\ TeV}}

\newcommand{\Lag}{{\mathcal L}}

\newcommand{\SUD}{{SU(2)_{\rm D}}}

\newcommand{\Z}{{\mathbb Z}}

\newcommand{\lhapdf}{{\sc\small LHAPDF6}}
\newcommand{\mg}{{\sc \texttt{MG5\_aMC}}}

\newcommand{\pythia}{{\sc \texttt{Pythia}}}

\newcommand{\be}{\begin{equation}}
\newcommand{\ee}{\end{equation}}
\def\bsp#1\esp{\begin{split}#1\end{split}}
\def\bpm{\begin{pmatrix}}
\def\epm{\end{pmatrix}}

\newcommand{\com}[1]{\iffalse #1 \fi}

\usepackage{xcolor}

\usepackage[normalem]{ulem}
\crefname{equation}{Eq.}{Eqs.}
\crefname{figure}{Fig.}{Figs.}
\crefname{tabular}{Tab.}{Tabs.}

\usepackage[T1]{fontenc} % if needed

\newcommand{\Br}{\operatorname{Br}}

\begin{document}

\title{\boldmath Novel Multilepton Signatures from  the Fermionic Portal to Vector Dark Matter}

\author{Alexander Belyaev }
\email{a.belyaev@soton.ac.uk}
\affiliation{School of Physics and Astronomy, University of Southampton, Highfield, Southampton SO17 1BJ, UK}
\affiliation{Particle Physics Department, Rutherford Appleton Laboratory, Chilton, Didcot, Oxon OX11 0QX, UK}

\author{Manimala Chakraborti}
\email{mani.chakraborti@gmail.com}
\affiliation{School of Physics and Astronomy, University of Southampton, Highfield, Southampton SO17 1BJ, UK}

\author{Claire Shepherd-Themistocleous}
\email{C.H.Shepherd-Themistocleous@rl.ac.uk}
\affiliation{Particle Physics Department, Rutherford Appleton Laboratory, Chilton, Didcot, Oxon OX11 0QX, UK}

\author{Chang-Yuan Yao}
\email{c.yao@soton.ac.uk}
\affiliation{School of Physics and Astronomy, University of Southampton, Highfield, Southampton SO17 1BJ, UK}

\begin{abstract}
We perform a collider study of a novel multilepton signature arising
from pair production of heavy vector-like leptons followed by cascade
decays through a dark sector. In the muonic realisation of the
Fermionic Portal to Vector Dark Matter, this process can lead to final
states with four, six, eight, or ten visible muons, depending on the
dark-sector spectrum and branching pattern. We identify the six-muon
channel as the most powerful target: it remains sizeable over a broad
region of parameter space, while being less rate-suppressed than the
higher-multiplicity channels and much cleaner and more reconstructable
than the four-muon final state. The signal arises from Drell--Yan pair
production of vector-like muons, $pp\to\mu'^{+}\mu'^{-}$, followed by decays
through the dark vector $V'$ and the dark scalar $H_D$. The six-muon
final state receives contributions from symmetric decay topologies in
which each $\mu'$ yields three visible muons, and from the asymmetric
topology in which one chain yields one muon and the other yields five.
The six-lepton signature from vector-like-lepton pair production has
not previously been explored at the LHC. We therefore develop a
topology-based reconstruction which exploits the repeated dimuon,
trimuon, and five-muon resonance structure of the signal. We simulate
signal and Standard Model backgrounds at detector level, including a
dedicated treatment of rare heavy-flavour muons. The resulting
background after the six-muon selection and topology reconstruction is
negligible. Existing Run-2 multilepton searches already constrain part
of the low-mass parameter space, but they do not exploit the repeated
resonance structure of the signal. A dedicated six-muon search can
substantially extend the reach. At the HL-LHC, the proposed analysis
can probe vector-like muon masses up to about $1.9$ TeV for favourable
spectra.
\end{abstract}

\maketitle
\newpage
\tableofcontents
\newpage

%%%%%%%%%%%%%%%%%%%%%%%%%
%%%%%%%%%%%%%%%%%%%%%%%%

\section{Introduction}
\label{sec:intro}

Understanding the particle nature of Dark Matter (DM) remains one of the
central open problems in particle physics and cosmology. Astrophysical and
cosmological observations provide overwhelming evidence for a non-luminous
matter component, while measurements of the Cosmic Microwave Background (CMB)
determine its abundance with high precision~\cite{Planck:2018vyg}. Yet the
microscopic properties of DM remain unknown: its spin, mass scale, stability
mechanism, possible dark-sector structure, and non-gravitational interactions
with the Standard Model (SM) are all still open questions. This makes DM one
of the strongest motivations for physics beyond the Standard Model (BSM).

A particularly important collider possibility is that the dark sector is not
accessed through a single mediator resonance, but through the production and
decay of new fermionic portal states~\cite{Belyaev:2022shr,Belyaev:2022zjx}.
Pair-produced vector-like fermions can give rise to multi-step cascades in
which each heavy fermion decay chain contains several visible SM
particles~\cite{Falkowski:2013jya,Bissmann:2020lge,Dermisek:2022xal}.
This naturally leads to six-fermion final states, with a characteristic
two-branch structure inherited from the pair-production process. Such
signatures can arise in a broad class of BSM models with vector-like fermion
portals, but they have not been systematically exploited as reconstructable
LHC topologies. In this work we study the clean muonic realisation of this
idea, where the final state can contain six visible muons and the underlying
cascade can be reconstructed through repeated dimuon and trimuon resonance
structures.
A particularly well-motivated possibility is that DM is not an isolated
particle, but part of a larger dark gauge sector. In
Refs.~\cite{Belyaev:2022shr,Belyaev:2022zjx}, the Fermionic Portal to Vector
Dark Matter (FPVDM) framework was introduced, in which the DM candidate is a
massive gauge boson of a non-Abelian dark symmetry and the connection to the
SM is mediated by new vector-like fermions. This construction differs from
ordinary Higgs-portal vector-DM scenarios because the dominant link between
the visible and dark sectors is fermionic rather than scalar. As a result,
the same structure that stabilises the vector DM state can also give rise to
distinctive collider signatures involving the vector-like fermion partners.

The muon realisation of this framework, the Muonic Portal to Vector Dark
Matter (MPVDM), was studied in Ref.~\cite{Belyaev:2025cgf}. In this model the
new vector-like fermion doublet,
\[
\Psi =
\begin{pmatrix}
\mu_D \\
\mu'
\end{pmatrix},
\]
couples the dark sector to the SM muon. The spectrum contains a stable dark
vector state, additional dark gauge and scalar states, and heavy vector-like
muon partners. The MPVDM model was shown in Ref.~\cite{Belyaev:2025cgf} to
provide a correlated framework for DM phenomenology, precision muon
observables, and collider searches. In particular, that study identified
striking multilepton signatures as a promising consequence of the model.
The present paper develops this observation into a dedicated collider study of
the six-muon topology and its reconstruction.

The key collider process considered here is pair production of the heavy
vector-like muon partner,
\[
pp \to \mu'^{+}\mu'^{-},
\]
followed by cascade decays through the new dark-sector states. Depending on
the decay chain, this production mode can lead to final states with four,
six, eight, or ten muons. Among these possibilities, we find that the $6\mu$
final state plays a special role. It is sufficiently rare in the SM to
be essentially background-free after topology reconstruction, but it is less
rate-suppressed and more generic than the eight- and ten-muon channels. It
also retains enough internal resonance structure to reconstruct the
intermediate dark states and the parent vector-like lepton. In this sense,
the $6\mu$ channel provides the best balance between cleanliness, rate, and
model coverage.

There is an important complementarity with the previously studied
$2\mu+\met$ channel in Ref.~\cite{Belyaev:2025cgf}. When the mass splitting
between $\mu'$ and $\mu_D$ is large and $V_D$ is light, the decay
\[
\mu'\to \mu_D V_D
\]
can dominate. In that regime, $\mu'^{+}\mu'^{-}$ production mainly feeds
$2\mu+\met$ final states after $\mu_D\to\mu V_D$, and the phenomenology is
closely aligned with direct $\mu_D^+\mu_D^-$ production. Since $\mu_D$ is
lighter than $\mu'$, the direct $\mu_D^+\mu_D^-$ cross section is then
typically much larger than the $\mu'^{+}\mu'^{-}$ cross section, which falls
rapidly with the heavy-lepton mass. This regime was analysed in
Ref.~\cite{Belyaev:2025cgf}. The present work focuses on the complementary
region in which $\mu'$ and $\mu_D$ are relatively close in mass, so that
$\mu'\to\mu_DV_D$ does not dominate, and the multimuon cascades from
$\mu'\to\mu V'$ and $\mu'\to\mu H_D$ become the leading discovery target.

The goal of this paper is therefore twofold. First, we identify the $6\mu$
signature as the most promising multilepton probe of the MPVDM model and
study its behaviour over the full parameter space relevant to this topology.
More generally, we use this model to demonstrate how six-lepton final states
from vector-like fermion pair production can reveal the structure of the
underlying BSM cascade, rather than appearing only as inclusive multilepton
excesses.
Secondly, we develop an optimised search and reconstruction strategy for this
final state. We perform detector-level simulations of both signal and
background, design event selections adapted to the characteristic cascade
topology, and reconstruct the masses of the heavy vector-like muon, the
intermediate dark gauge boson $V'$, and the dark scalar $H_D$. This mass
reconstruction is an essential part of the analysis: the signal is not merely
an excess of leptons, but a kinematically structured BSM event topology.

We deliberately formulate the analysis as an explicit topology-based
reconstruction rather than as a black-box classifier. By this we mean that
the event selection is not based only on an opaque signal-background score,
but on explicit assignments of reconstructed muons to the decay products of
the cascade. The reason is that the essential problem is not only to separate
signal from background, but to identify the cascade structure, resolve the
combinatorics of the six-muon final state, and reconstruct the intermediate
$V'$, $H_D$, and $\mu'$ states. Machine-learning tools, such as neural
networks or boosted decision trees, could be useful in a future optimisation
of the analysis, but the present strategy is designed to keep the physics
content of the reconstruction explicit.

By applying this explicit reconstruction strategy,
we find a gap between the sensitivity of existing non-dedicated
searches and the reach of a dedicated $6\mu$ analysis. Current LHC searches,
which are not optimised for this topology, constrain only part of the
low-mass parameter space. A dedicated analysis based on the repeated dimuon
and trimuon resonance structure substantially improves the sensitivity in the
same luminosity regime. At the HL-LHC, the optimised strategy can probe
vector-like muon masses up to about $1.9~\TeV$ for favourable spectra. This
illustrates the physics gain from treating the six-muon final state as a
reconstructable cascade topology rather than as a generic multilepton event.

The implications of this study extend beyond the specific MPVDM benchmark.
Multilepton searches at the LHC are usually framed in terms of supersymmetry,
heavy neutral leptons, vector-like leptons, or resonant leptonic scenarios,
as illustrated by existing ATLAS and CMS searches in multilepton final
states~\cite{CMS:2017moi,ATLAS:2019lng,CMS:2018jxx,CMS:2024ake,ATLAS:2023sbu,ATLAS:2022pbd}.
The MPVDM model provides a different target: a fermionic portal to a dark
gauge sector, with a characteristic cascade structure and reconstructable
intermediate states. The $6\mu$ signature therefore offers a clean and powerful
way to test a class of BSM theories that would otherwise remain only weakly
constrained by standard searches.

The paper is organised as follows. In Sec.~\ref{sec:model} we summarise the
MPVDM model and the particle spectrum relevant for the collider analysis. In
Sec.~\ref{sec:signatures} we discuss the multimuon signatures arising from
vector-like muon pair production, compare the relative rates of the four-,
six-, eight-, and ten-muon final states, and identify the $6\mu$ channel as
the main target of this study. We then introduce the benchmark scenarios used
for the detector-level study. The event simulation, background
treatment, topology-based reconstruction, selection strategy, and mass reach
are presented in Sec.~\ref{sec:collider_analysis}. The resulting current and
projected sensitivities are presented before we conclude in
Sec.~\ref{sec:conclusion}.

%%%%%%%%%%%%%%%%%%%%%%%%%%%%%%
%%%%%%%%%%%%%%%%%%%%%%%%%%%%%
%%%%%%%%%%%%%%%%%%%%%%%%%%%%%%%%%%%
%%%%%%%%%%%%%%%%%%%%%%%%%%%%%%%%%%%

\section{The MPVDM model}

\label{sec:model}

In this section we briefly summarise the ingredients of the Muonic Portal to Vector Dark Matter (MPVDM) model that are relevant for the collider analysis. The full construction and its dark-matter, precision, and cosmological phenomenology were discussed in Refs.~\cite{Belyaev:2022shr,Belyaev:2022zjx,Belyaev:2025cgf}. Here we focus on the particle content, mass spectrum, and interactions that lead to the multimuon signatures studied in this paper.

The MPVDM model is a muonic realisation of the Fermionic Portal to Vector Dark Matter (FPVDM) framework. The dark sector is based on a non-Abelian gauge symmetry $SU(2)_D$, broken by a dark scalar doublet $\Phi_D$. The corresponding dark gauge bosons acquire masses through the dark-sector Higgs mechanism. In contrast to ordinary Higgs-portal vector-DM models, the dominant connection between the dark and visible sectors is provided by new vector-like fermions rather than by the scalar portal. In the present work the relevant vector-like fermion is a muon partner, so that the dark sector communicates primarily with the SM muon.

The model contains a dark scalar doublet,
\[
\Phi_D =
\begin{pmatrix}
\varphi^0_{D+1/2} \\
\varphi^0_{D-1/2}
\end{pmatrix},
\]
a vector-like fermion doublet under $SU(2)_D$,
\[
\Psi =
\begin{pmatrix}
\psi_{\mu_D} \\
\psi_{\mu'}
\end{pmatrix},
\]
and the dark gauge bosons
\[
V^D_\mu =
\begin{pmatrix}
V^0_{D+\mu} \\
V^0_{D0\mu} \\
V^0_{D-\mu}
\end{pmatrix}.
\]
Following the notation of Refs.~\cite{Belyaev:2022shr,Belyaev:2022zjx,Belyaev:2025cgf}, we denote the $\Z_2$-odd dark vector state by $V_D$ and the neutral $\Z_2$-even dark vector by $V'$. The $\Z_2$ symmetry stabilises the lightest $\Z_2$-odd state, which is the vector DM candidate in the parameter region considered here.

The quantum numbers of the new particles are summarised in \cref{tab:particlesQN}. The relevant residual charge is
\[
Q_D = T^3_D + Y_D,
\]
and the unbroken discrete symmetry is
\[
\Z_2 = (-1)^{Q_D}.
\]
With the assignments $Y_D=1/2$ for the dark scalar and fermion doublets and $Y_D=0$ for the vector triplet, this symmetry remains after dark symmetry breaking and ensures the stability of the lightest $\Z_2$-odd state.

\setlength{\tabcolsep}{3pt}
\setlength{\arraycolsep}{0pt}
\begin{table}[htbp]
\centering
\begin{tabular}{c|cc|cc||c|r}
\toprule
 & $SU(2)_L$ & $U(1)_Y$ & $\SUD$ & $U(1)_{Y_D}$ & $\Z_2$ & $Q_D$\\
\midrule
$\Phi_{D}=\left(\begin{array}{c} \varphi^0_{D+ \frac{1}{2} } \\ \varphi^0_{D-\frac{1}{2} } \end{array}\right)$ & $\mathbf{1}$ & $0$ & $\mathbf{2}$
& $\frac{1}{2}$
& $\begin{array}{c} - \\ + \end{array}$
& $\begin{array}{r} +1 \\ 0 \end{array}$
\\[2pt]
\midrule
$\Psi=\left(\begin{array}{c} \psi_{\mu_D} \\ \psi_{\mu'} \end{array}\right)$ & $\mathbf{1}$ & $-1$ & $\mathbf{2}$ & $\frac{1}{2}$ & $\begin{array}{c} - \\ + \end{array}$ & $\begin{array}{r} +1 \\ 0 \end{array}$\\[2pt]
\midrule
$V^D_{\mu}=\left(\begin{array}{c} V^0_{D+\mu} \\ V^0_{D0\mu} \\ V^0_{D-\mu} \end{array}\right)$ & $\mathbf{1}$ & $0$ & $\mathbf{3}$
& $0$
& $\begin{array}{c} - \\ + \\ - \end{array}$
& $\begin{array}{r} +1 \\ 0 \\ -1 \end{array}$
\\[2pt]
\bottomrule
\end{tabular}
\caption{\label{tab:particlesQN}Quantum numbers of the new MPVDM particles under the EW and dark-sector gauge groups.}
\end{table}

The relevant terms in the Lagrangian are
\begin{eqnarray}
\Lag &\supset& - \frac{1}{4} (V_{\mu\nu}^i)^2|_{B,W^i,V^i_D}
+ \bar{f}^{\rm SM} i \slashed{D} f^{\rm SM}
+ \bar{\Psi} i \slashed{D} \Psi
+ | D_\mu \Phi_H |^2
+ | D_\mu \Phi_D |^2
- V(\Phi_H, \Phi_D)
\nonumber \\
&-&
\left(
y \bar{f}^{\rm SM}_L \Phi_H f^{\rm SM}_R
+ y^\prime \bar{\Psi}_L \Phi_D \mu_R
+ h.c.
\right)
- M_\Psi \bar{\Psi} \Psi \;,
\label{eq:Lagrangian_dark_sector}
\end{eqnarray}
where the scalar potential is
\begin{equation}
V(\Phi_H,\Phi_D) =
- \mu^2 \Phi_H^\dagger \Phi_H
- \mu_D^2 \Phi_D^\dagger \Phi_D
+ \lambda (\Phi_H^\dagger \Phi_H)^2
+ \lambda_D (\Phi_D^\dagger \Phi_D)^2
+ \lambda_{HD}(\Phi_H^\dagger \Phi_H)(\Phi_D^\dagger \Phi_D)\;.
\label{eq:scalarpotential}
\end{equation}
In this work we focus on the fermionic portal and set the scalar portal coupling to be negligible,
\[
\lambda_{HD} \simeq 0,
\]
or equivalently take the scalar mixing angle to be zero at tree level. This choice keeps the SM-like Higgs phenomenology essentially unchanged and isolates the collider effects associated with the vector-like muon and dark gauge sector.

After EW and dark symmetry breaking, the neutral component of $\Phi_D$ acquires a VEV $v_D$, while the SM Higgs doublet has the usual VEV $v$. The dark gauge bosons have the common tree-level mass
\begin{equation}
m_V \equiv m_{V'} = m_{V_D} = \frac{g_D v_D}{2}\;.
\label{eq:VPmass}
\end{equation}
Loop corrections split the masses of $V_D$ and $V'$. In the parameter region relevant for this study, the mass difference is approximately
\begin{align}
m_{V_D}-m_{V'} =
\frac{1}{3}
\frac{g_D^2m_{\mu'}^2}{32 \pi^2 m_{V_D}}
\left(
\frac{m_{\mu'}^2-m_{\mu_D}^2}{m_{\mu'}^2}
\right)^2\;.
\label{eq:simple_mass_splitting_2}
\end{align}
This splitting is important for the dark-matter phenomenology of the model, but in the collider analysis below the relevant point is that the spectrum contains both the stable dark vector $V_D$ and the unstable dark vector $V'$.

The Yukawa interaction in \cref{eq:Lagrangian_dark_sector} mixes the SM muon with the $\Z_2$-even vector-like fermion interaction eigenstate $\psi_{\mu'}$, yielding the physical mass eigenstates denoted by $\mu$ (the SM muon) and $\mu'$ (the heavy muon partner). The $\Z_2$-odd fermion $\psi_{\mu_D}$ does not mix with the SM muon, and its mass eigenstate is denoted by $\mu_D$. The physical fermion masses are
\begin{equation}
m_{\mu_D}=M_\Psi
\end{equation}
and
\begin{equation}
m_{\mu,\mu'}^2=
\frac{1}{4}
\left[
y^2 v^2 + y^{\prime 2} v_D^2 + 2 M_\Psi^2
\mp
\sqrt{
\left(y^2 v^2 + y^{\prime 2} v_D^2 + 2 M_\Psi^2\right)^2
-8y^2v^2M_\Psi^2
}
\right]\;.
\label{eq:fermionic_masses}
\end{equation}
The resulting hierarchy is
\begin{equation}
m_\mu < m_{\mu_D} \leq m_{\mu'}\;.
\label{eq:fermionic_mass_hierarchy}
\end{equation}
The mixing angles between the SM muon and the vector-like muon are
\begin{equation}
\sin\theta_{R} =
\sqrt{
\frac{m_{\mu'}^2 - m_{\mu_D}^2}
     {m_{\mu'}^2 - m_\mu^2}
}
\quad\text{and}\quad
\sin\theta_{L} =
\frac{m_\mu}{m_{\mu_D}}\sin\theta_{R}\;.
\label{eq:fermion_mixing_angles}
\end{equation}
Equivalently, the Yukawa couplings can be written in terms of the physical masses and VEVs as
\begin{equation}
y =
\sqrt{2}\frac{m_\mu m_{\mu'}}{m_{\mu_D}v},
\qquad
y^\prime =
\sqrt{2}
\frac{
\sqrt{(m_{\mu'}^2-m_{\mu_D}^2)(m_{\mu_D}^2-m_\mu^2)}
}
{m_{\mu_D}v_D}\;.
\label{eq:yukawas}
\end{equation}
The limit $m_{\mu'}=m_{\mu_D}$ corresponds to $y^\prime=0$ and vanishing fermion mixing, so that the new fermion sector decouples from the SM muon.

The scalar sector contains the SM-like Higgs boson $h$ and a dark scalar state, denoted here by $H_D$. In the limit $\lambda_{HD}=0$, these states do not mix at tree level and their masses are controlled separately by the SM and dark scalar potentials. The dark scalar mass can therefore be treated as an independent parameter in the collider analysis.

The five independent parameters used in this study are chosen as
\begin{equation}
g_D,\qquad
m_{V_D},\qquad
m_{H_D},\qquad
m_{\mu_D},\qquad
m_{\mu'} .
\label{eq:pars}
\end{equation}
The remaining quantities are then fixed by
\begin{eqnarray}
{\setlength{\arraycolsep}{5pt}
\begin{array}{ccc}
v = 2 \dfrac{m_W}{g}, &
v_D = 2 \dfrac{m_{V_D}}{g_D}, \\[6pt]
\lambda = \dfrac{g^2m_h^2}{8m_W^2}, &
\lambda_D = \dfrac{g_D^2m_{H_D}^2}{8m_{V_D}^2}, &
\lambda_{HD}=0, \\[8pt]
y = \dfrac{g m_\mu}{\sqrt{2}m_W}\dfrac{m_{\mu'}}{m_{\mu_D}}, &
y^\prime =
\dfrac{
g_D\sqrt{(m_{\mu'}^2-m_{\mu_D}^2)(m_{\mu_D}^2-m_\mu^2)}
}
{\sqrt{2}m_{\mu_D}m_{V_D}} .
\end{array}
}
\label{eq:dependent_parameters}
\end{eqnarray}

The collider phenomenology studied in this paper is driven by pair production of the heavy vector-like muons,
\begin{equation}
pp \to \mu'^{+}\mu'^{-}.
\label{eq:mupairproduction}
\end{equation}
Since $\mu'$ carries the same EW quantum numbers as the SM right-handed muon, this process proceeds through the standard Drell-Yan production mechanism. The subsequent decays of the produced $\mu'$ states are controlled by the dark-sector interactions in \cref{eq:Lagrangian_dark_sector} and can generate high-multiplicity muon final states accompanied by missing transverse momentum from stable $V_D$ particles. The detailed cascade structure, including the origin of the characteristic six-muon signature, is discussed in the next section.
%%%%%%%%%%%%%%%%%%%%%%%%%%%%%%%%%%%%%%%%%%%%%%%%%%%%%%
%%%%%%%%%%%%%%%%%%%%%%%%%%%%%%%%%%%%%%%%%%%%%%%%%%%%%%
%%%%%%%%%%%%%%%%%%%%%%%%%%%%%%%%%%%
%%%%%%%%%%%%%%%%%%%%%%%%%%%%%%%%%%%
\section{Multimuon collider signatures}

\label{sec:signatures}

The characteristic collider phenomenology of the MPVDM model is driven by pair production of the heavy vector-like muon,
\begin{equation}
pp\to \mu'^{+}\mu'^{-}.
\label{eq:mupair_signatures}
\end{equation}
Since $\mu'$ has the same EW quantum numbers as the SM right-handed muon, the production process proceeds dominantly through the standard Drell--Yan mechanism mediated by an $s$-channel $\gamma/Z$. The production cross section therefore depends mainly on the heavy-lepton mass $m_{\mu'}$, while the subsequent decay pattern is controlled by the dark-sector spectrum and couplings.

The related process
\begin{equation}
pp\to \mu_D^+\mu_D^-,
\end{equation}
followed by $\mu_D\to\mu V_D$, gives predominantly a
$2\mu+\met$ signature and was studied in
Ref.~\cite{Belyaev:2025cgf}. Since $\mu_D$ is lighter than $\mu'$, its
Drell--Yan production rate is generally larger. The present work instead
focuses on $\mu'^{+}\mu'^{-}$ production. Although its production rate can
be smaller, the decay $\mu'\to\mu V'$ has substantially more available
phase space than $\mu'\to\mu_DV_D$, producing a more energetic muon, while
the subsequent decays of $V'$ and $H_D$ generate the distinctive multimuon
cascades studied below.
Moreover, the $\mu'^{+}\mu'^{-}$ signal is straightforward to trigger on,
since the two primary muons from the $\mu'$ decays are typically energetic
and are always present in the multimuon channels considered here.

Depending on the decay chains of the two $\mu'$ states and of the intermediate $V'$ and $H_D$ resonances, the visible final state can contain two, four, six, eight, or ten muons, possibly accompanied by missing transverse momentum from stable $V_D$ particles. The higher-multiplicity final states with four or more muons are the main subject of this paper.

The relevant decay modes of the heavy vector-like muon are
\begin{equation}
\mu' \to \mu V',
\qquad
\mu' \to \mu H_D,
\qquad
\mu' \to \mu_D V_D .
\label{eq:mup_decay_modes}
\end{equation}
The first two modes generate visible multimuon cascades through the subsequent decays
\begin{equation}
V'\to \mu^+\mu^-,
\label{eq:Vp_mumu}
\end{equation}
and
\begin{equation}
H_D\to \mu^+\mu^-,
\qquad
H_D\to V'V'\to 4\mu,
\qquad
H_D\to V_DV_D^* .
\label{eq:HD_decay_modes_signature}
\end{equation}
The decay $H_D\to V_DV_D^*$ produces missing transverse momentum and reduces the visible muon multiplicity, while $H_D\to V'V'\to4\mu$ gives a distinctive four-muon subsystem, which can be boosted when $H_D$ is produced with large momentum.

\begin{figure}[htb]
\centering
    \begin{tabular}{c c}
        \includegraphics[trim={0cm 0cm 0 0cm},clip,width=0.76\textwidth]{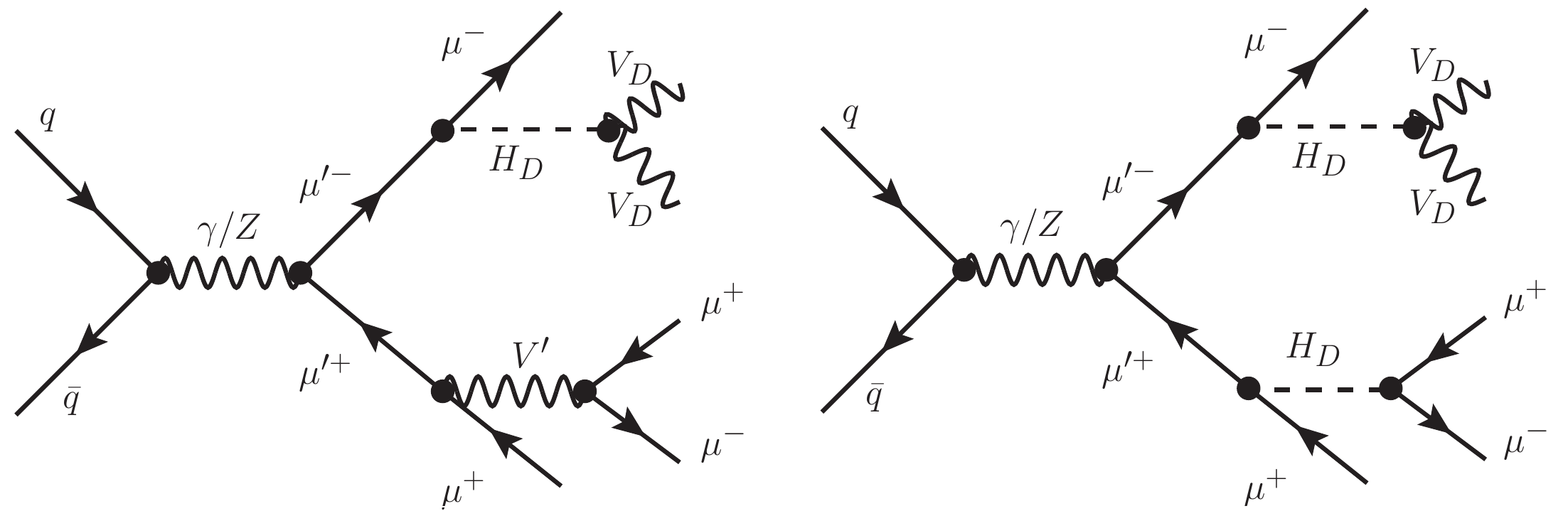} &
        \\
        (A) \hspace*{5cm} (B)&\\
     \end{tabular}
\caption{\label{fig:4mu-diags}
Representative Feynman diagrams leading to four visible muons in $\mu'^{+}\mu'^{-}$ production. These topologies arise when one decay chain produces one visible muon while the other produces three visible muons.}
\end{figure}

\begin{figure}[htb]
\centering
    \begin{tabular}{c c}
        \includegraphics[trim={3cm 16cm 0 0cm},clip,width=0.76\textwidth]{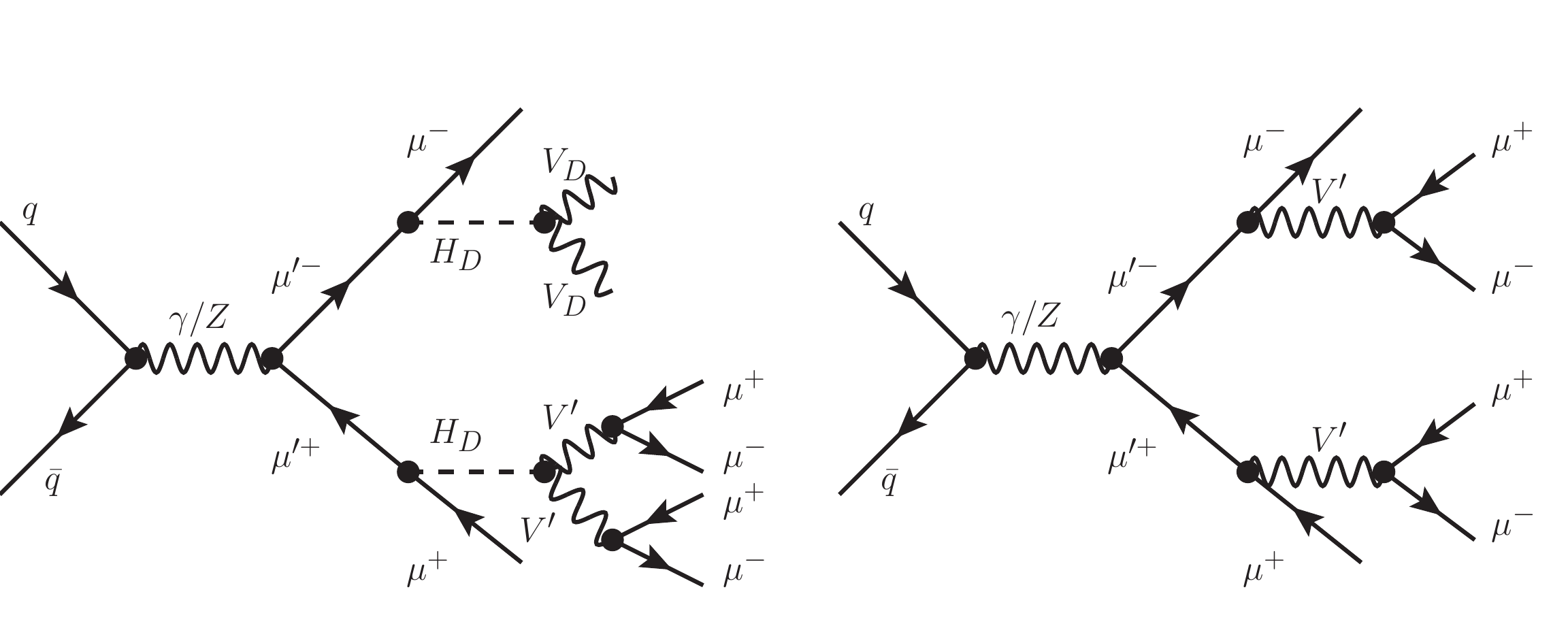} &\\
        (A) \hspace*{5cm} (B)&\\
        \includegraphics[trim={3cm 3cm 0 15cm},clip,width=0.76\textwidth]{fig/multi-leptons-diag-6.pdf} &\\
        (C) \hspace*{5cm} (D)&\\
        \end{tabular}
\caption{\label{fig:6mu-diags}
Representative Feynman diagrams leading to six visible muons. The $6\mu$ final state receives contributions both from asymmetric $(1+5)\mu$ topologies and from symmetric $(3+3)\mu$ topologies, making it the most robust and phenomenologically important multimuon channel.}
\end{figure}

\begin{figure}[htb]
\centering
    \begin{tabular}{c c}
        \includegraphics[trim={0cm 0cm 0 0cm},clip,width=0.76\textwidth]{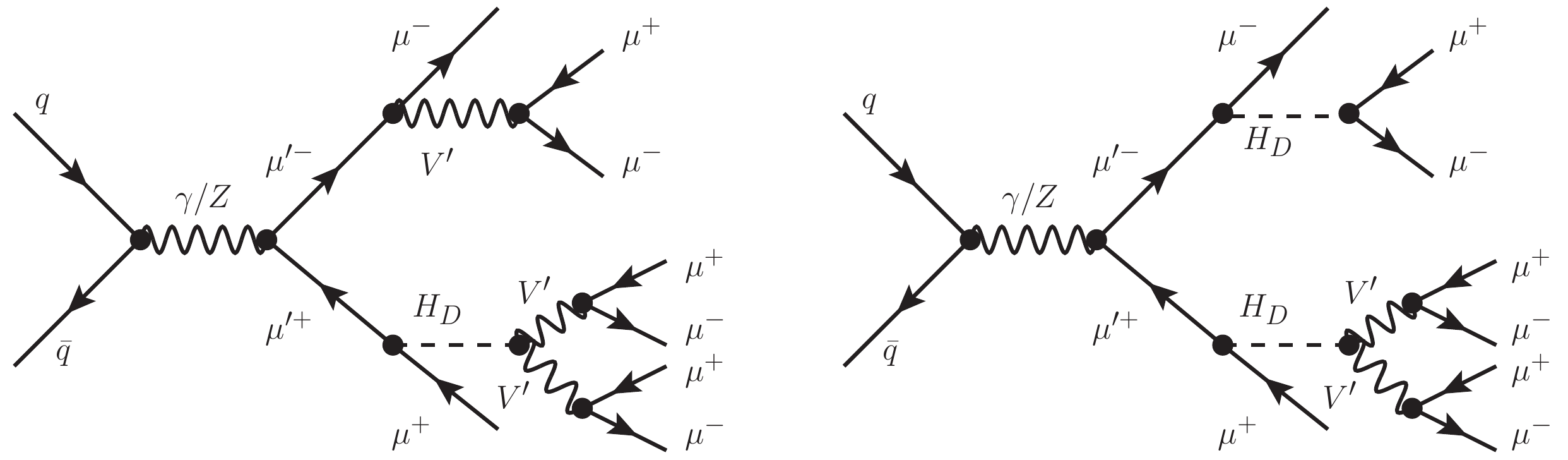} &
        \\
        (A) \hspace*{5cm} (B)&\\
    \end{tabular}
\caption{\label{fig:8mu-diags}
Representative Feynman diagrams leading to eight visible muons. These topologies involve one three-muon chain and one five-muon chain.}
\end{figure}

\begin{figure}[htb]
\centering
    \begin{tabular}{c c}
        \includegraphics[trim={0cm 0cm 0 0cm},clip,width=0.38\textwidth]{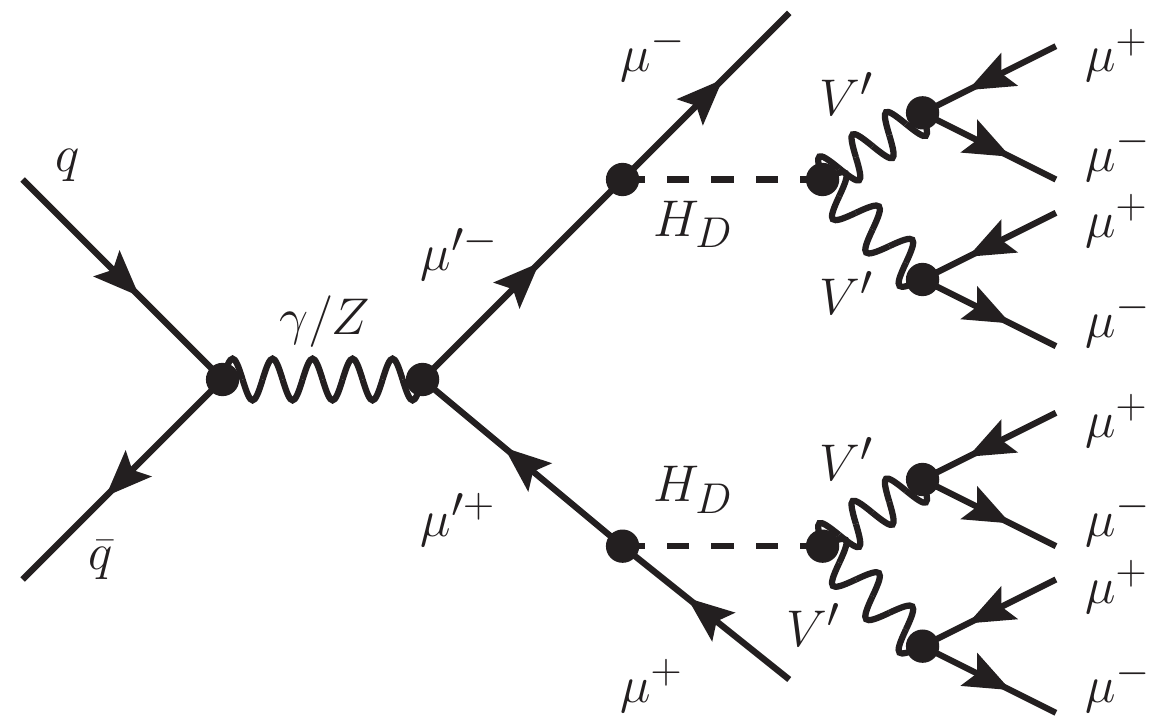} &\\
    \end{tabular}
\caption{\label{fig:10mu-diags}
Representative Feynman diagram leading to ten visible muons. This final state is extremely clean but rate-suppressed because both $\mu'$ decay chains must produce five visible muons.}
\end{figure}

It is useful to classify the possible signatures in terms of the number of visible muons produced in a single $\mu'$ decay chain. A one-muon chain can arise from
\begin{equation}
\mu'\to \mu H_D,
\qquad
H_D\to V_DV_D^* ,
\label{eq:one_mu_chain_HD}
\end{equation}
or from
\begin{equation}
\mu'\to \mu_D V_D,
\qquad
\mu_D\to\mu V_D .
\label{eq:one_mu_chain_muD}
\end{equation}
A three-muon chain can arise either from
\begin{equation}
\mu'\to \mu V',
\qquad
V'\to \mu^+\mu^-,
\label{eq:three_mu_chain_Vp}
\end{equation}
or from
\begin{equation}
\mu'\to \mu H_D,
\qquad
H_D\to \mu^+\mu^- .
\label{eq:three_mu_chain_HD}
\end{equation}
Finally, a five-muon chain is generated by
\begin{equation}
\mu'\to \mu H_D,
\qquad
H_D\to V'V'\to 4\mu .
\label{eq:five_mu_chain}
\end{equation}
Combining the two decay chains in $\mu'^{+}\mu'^{-}$ production then gives the visible muon multiplicities
\begin{equation}
(1+1)\mu=2\mu,\qquad
(1+3)\mu=4\mu,\qquad
(1+5)\mu=(3+3)\mu=6\mu,
\label{eq:visible_muon_multiplicities}
\end{equation}
and
\begin{equation}
(3+5)\mu=8\mu,\qquad
(5+5)\mu=10\mu .
\end{equation}
The representative Feynman diagrams for the four-, six-, eight-, and ten-muon final states are shown in
\cref{fig:4mu-diags,fig:6mu-diags,fig:8mu-diags,fig:10mu-diags}.

%%%%%%%%%%%%%%%%%%%%%%%%%%%%%%%%%%%
\subsection{Muon-multiplicity branching fractions}
\label{subsec:muon_multiplicity_branching}

To quantify the relative importance of the different pure-muon final states, we define the one-chain probabilities
$P_1$, $P_3$, and $P_5$, corresponding respectively to one, three, and five muons produced in a single $\mu'$ decay chain:
\begin{align}
P_1 &=
\Br(\mu'\to \mu H_D)\,
\Br(H_D\to V_DV_D^*)
+
\Br(\mu'\to \mu_D V_D)\,
\Br(\mu_D\to \mu V_D),
\label{eq:P1_definition}
\\
P_3 &=
\Br(\mu'\to \mu V')\,
\Br(V'\to\mu^+\mu^-)
+
\Br(\mu'\to \mu H_D)\,
\Br(H_D\to\mu^+\mu^-),
\label{eq:P3_definition}
\\
P_5 &=
\Br(\mu'\to \mu H_D)\,
\Br(H_D\to V'V')\,
\Br(V'\to\mu^+\mu^-)^2 .
\label{eq:P5_definition}
\end{align}

The corresponding branching fractions for the visible muon multiplicities in $\mu'^{+}\mu'^{-}$ events are then
\begin{equation}
\begin{aligned}
{\cal B}_{2\mu} &= P_1^2, &
{\cal B}_{4\mu} &= 2P_1P_3, &
{\cal B}_{6\mu} &= P_3^2 + 2P_1P_5, &
{\cal B}_{8\mu} &= 2P_3P_5, &
{\cal B}_{10\mu} &= P_5^2 .
\end{aligned}
\label{eq:multi_muon_branching_fractions}
\end{equation}

The sum of the pure-muon branching fractions is
\begin{equation}
{\cal B}_{2\mu}
+
{\cal B}_{4\mu}
+
{\cal B}_{6\mu}
+
{\cal B}_{8\mu}
+
{\cal B}_{10\mu}
=
(P_1+P_3+P_5)^2 .
\label{eq:pure_muon_sum}
\end{equation}
The one-chain probabilities $P_1$, $P_3$, and $P_5$ defined in
\cref{eq:P1_definition,eq:P3_definition,eq:P5_definition} saturate the
$\mu'$ decay width up to the EW decay modes induced by the small
$\mu$--$\mu'$ mixing,
\begin{equation}
\mu'\to Z\mu,
\qquad
\mu'\to W\nu_\mu .
\end{equation}
These channels are suppressed by at least five orders of magnitude and
contribute at most at the level of $\mathcal{O}(10^{-4})$ to the total
$\mu'$ branching ratio throughout the parameter space considered here.
Therefore,
\begin{equation}
P_1+P_3+P_5 \simeq 1,
\end{equation}
and hence
\begin{equation}
{\cal B}_{2\mu}
+
{\cal B}_{4\mu}
+
{\cal B}_{6\mu}
+
{\cal B}_{8\mu}
+
{\cal B}_{10\mu}
\simeq 1 .
\label{eq:branching_sum_rule}
\end{equation}

The relative importance of the four-, six-, eight-, and ten-muon channels is
quantified below after introducing the common Drell--Yan production rate.
This allows the branching-fraction patterns and the absolute multimuon rates
to be discussed in a consistent order.

%%%%%%%%%%%%%%%%%%%%%%%%%%%%%%%%
%%%%%%%%%%%%%%%%%%%%%%%%%%%%%%%

\subsection{Production rates for multimuon final states}
\label{subsec:production_rates}

\begin{figure}[htb]
\centering
\includegraphics[trim={0cm 0cm 0 0cm 0cm},clip,width=0.6\textwidth]{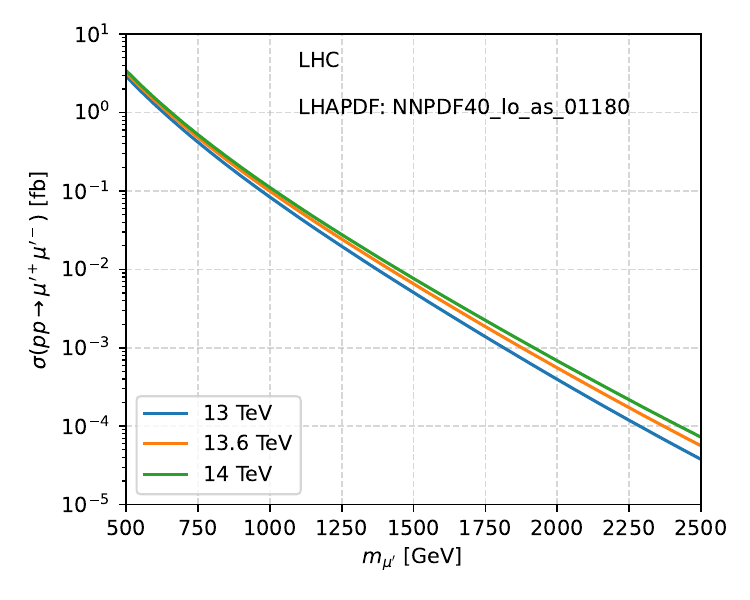}
\caption{\label{fig:xsec_pp2MM}
LO production cross section for $pp \to \mu'^{+}\mu'^{-}$ at
$\sqrt{s}=13,13.6$ and 14~TeV, shown as a function of $m_{\mu'}$. The calculation is
performed with \texttt{CalcHEP}, using the
\texttt{NNPDF40\_lo\_as\_01180} PDF set through \lhapdf{} and the scale choice
$Q=m_{\mu'}$. The production is dominated by the Drell--Yan $\gamma/Z$
channel and is therefore mainly controlled by the vector-like muon mass.}
\end{figure}

The inclusive production cross section for
\begin{equation}
pp\to \mu'^{+}\mu'^{-}
\end{equation}
is shown in \cref{fig:xsec_pp2MM} as a function of $m_{\mu'}$.
The result is evaluated at $\sqrt{s}=13.6~\TeV$ at LO with
\texttt{CalcHEP}~\cite{Belyaev:2012qa}, using the \texttt{NNPDF40\_lo\_as\_01180} PDF set accessed through
\lhapdf~\cite{Buckley:2014ana,NNPDF:2021njg} and
the QCD factorisation and renormalisation scale fixed to
\begin{equation}
Q=m_{\mu'} .
\end{equation}
Since $\mu'$ carries the EW quantum numbers of the SM right-handed muon,
the production proceeds through the Drell--Yan $\gamma/Z$ channel. The
cross section is therefore controlled mainly by $m_{\mu'}$ and decreases
rapidly as the vector-like muon becomes heavier.

The signal rate for a given visible muon multiplicity is obtained by
multiplying the inclusive production cross section by the corresponding
branching fraction,
\begin{equation}
\sigma_{N\mu}
=
\sigma(pp\to \mu'^{+}\mu'^{-})\,{\cal B}_{N\mu},
\qquad
N=4,6,8,10 .
\label{eq:sigma_Nmu}
\end{equation}
Thus, at fixed $m_{\mu'}$, the relative rates of the different multimuon
channels are determined by the branching fractions ${\cal B}_{N\mu}$,
while the absolute normalisation is fixed by the common Drell--Yan
production rate.

The sharp fall of the production cross section sets the relevant mass range
for the LHC study. In particular, even with the full HL-LHC luminosity of
$3~{\rm ab}^{-1}$, the sensitivity is expected to deteriorate once
$m_{\mu'}$ approaches the region where
\begin{equation}
\sigma(pp\to\mu'^{+}\mu'^{-}) \lesssim 10^{-3}~{\rm fb}.
\end{equation}
From \cref{fig:xsec_pp2MM}, this occurs slightly above the $2~\TeV$ scale. The
projected reach of the dedicated multimuon analysis is therefore expected
to lie around the  $m_{\mu'}\simeq 2~\TeV$, depending on the
branching fraction into the relevant six-muon topology and on the
reconstruction efficiency.

\begin{figure}[htb]
\centering
\includegraphics[trim={0cm 0cm 0 0cm 0cm},clip,width=0.99\textwidth]{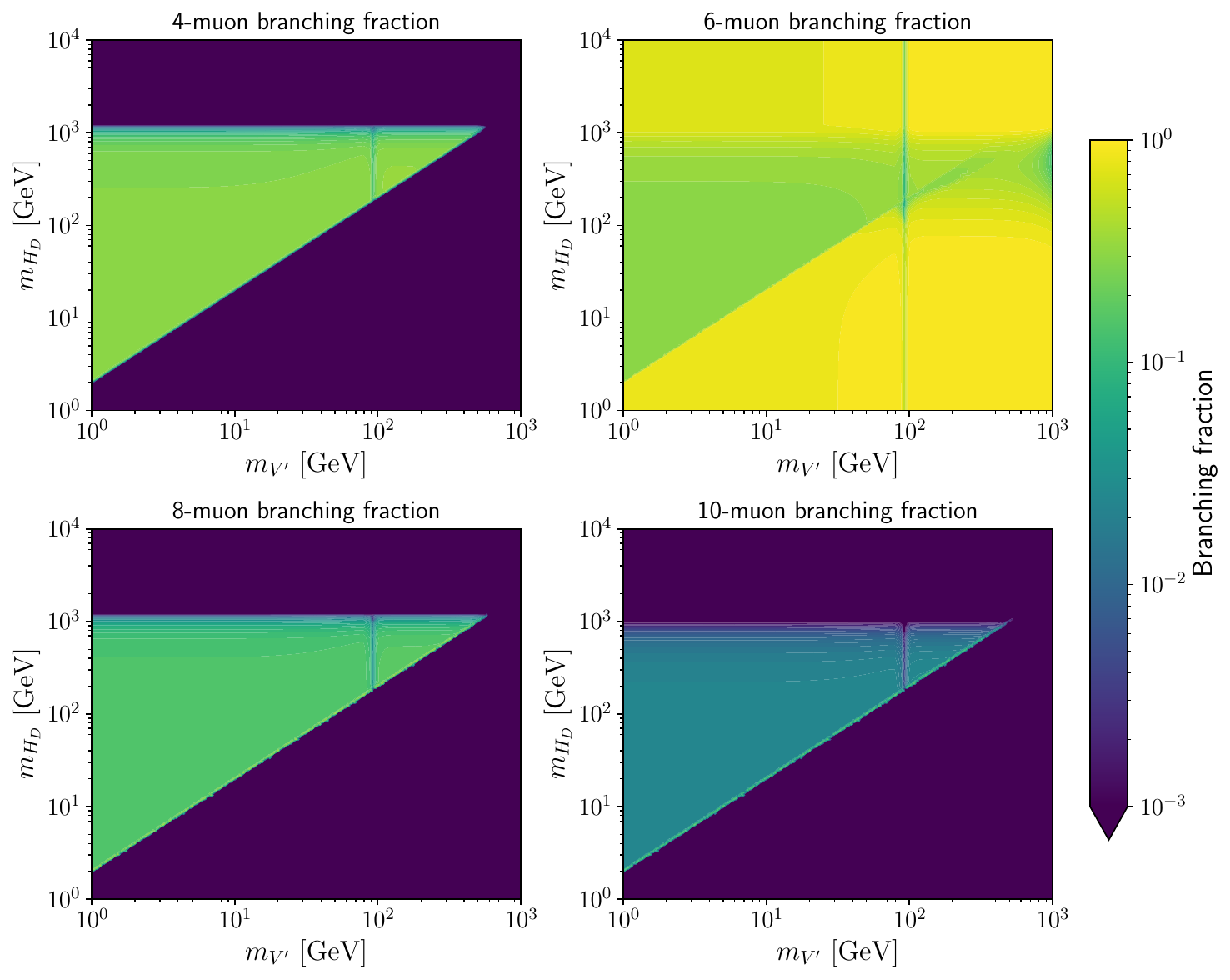}
\caption{\label{fig:multi_muon_fraction}
Branching fractions, equivalently relative rates at fixed
$\sigma(pp\to\mu'^{+}\mu'^{-})$, for visible four-, six-, eight-, and
ten-muon final states in $\mu'^{+}\mu'^{-}$ production, shown in the
$(m_{V'},m_{H_D})$ plane.
The branching ratios are evaluated for
$m_{\mu'}=1200~\GeV$, $m_{\mu_D}=1080~\GeV$, and
$g_D=10m_{V'}/m_{\mu'}$. The six-muon final state is sizeable over the
broadest region of the plane, making it the most generic multimuon topology
and the natural target for a dedicated reconstruction-based analysis. The
sharp feature near $m_{V'}\simeq m_Z$ is caused by resonant $V'$--$Z$
mixing, which enhances non-muonic $V'$ decays; this region is independently
strongly constrained by LEP measurements of the $Z$ properties. The feature
near $m_{H_D}=2m_{V'}$ marks the opening of the on-shell
$H_D\to V'V'$ channel, which enhances the eight- and ten-muon rates.}
\end{figure}

The branching fractions for the four-, six-, eight-, and ten-muon final
states are shown in \cref{fig:multi_muon_fraction} in the
$(m_{V'},m_{H_D})$ plane. The plot illustrates the complementarity of the
multimuon channels: the four-muon mode can be sizeable but is less distinctive,
whereas the eight- and ten-muon modes are cleaner but increasingly
rate-suppressed. The six-muon mode remains sizeable over the broadest region
of the plane.
The same parameter choice is also used for the
six-muon cross sections and the dark-scalar width presented in
\cref{fig:xsec_6mu_plane} and discussed below:
\begin{equation}
m_{\mu'}=1200~\GeV,
\qquad
m_{\mu_D}=0.9\,m_{\mu'}=1080~\GeV,
\qquad
g_D=10\,\frac{m_{V'}}{m_{\mu'}} .
\label{eq:scan_parameters}
\end{equation}
The $10\%$ splitting between $\mu'$ and $\mu_D$ is chosen as an intermediate
regime. A much smaller splitting would suppress the Yukawa coupling $y'$ and
could make $\mu'$ long-lived, which is not the phenomenology considered
here. Conversely, a substantially larger splitting would require a larger
$y'$ and could lead to a loss of perturbative control and an excessively
large $\mu'$ width. The adopted value therefore gives prompt $\mu'$ decays
while keeping the fermionic portal coupling and the heavy-lepton width under
perturbative control.

The dependence $g_D\propto m_{V'}$ follows the relic-density-compatible
pattern identified in Ref.~\cite{Belyaev:2025cgf}. In addition, $g_D$ is
chosen sufficiently large that the decays of $V'$ and $H_D$ remain prompt
on detector scales. Although their width-to-mass ratios can be small, the
corresponding decay lengths for the benchmark scenarios are still far below
the spatial resolution relevant for displaced-vertex searches. For
substantially smaller values of $g_D$, displaced or long-lived-particle
signatures could arise. Such signatures require a qualitatively different
experimental strategy and are not considered in the present analysis; they
will be studied separately.

The chosen relation also keeps the dark symmetry-breaking scale fixed:
\begin{equation}
v_D=\frac{2m_{V'}}{g_D}=\frac{m_{\mu'}}{5}.
\end{equation}
This prescription allows the dependence on $m_{V'}$ and $m_{H_D}$ to be
studied without simultaneously varying the dark-sector symmetry-breaking
scale.

At fixed $m_{\mu'}$, the maps in \cref{fig:multi_muon_fraction} also
represent the relative pattern of the cross sections
$\sigma_{N\mu}$ through \cref{eq:sigma_Nmu}. The key observation is that
the six-muon final state is the most generic multimuon topology in the
parameter space shown. Its branching fraction remains sizeable over the
broadest region of the $(m_{V'},m_{H_D})$ plane. By contrast, the four-muon final state arises from the $(1+3)\mu$
combination, while the eight- and ten-muon channels require at least one,
and in the ten-muon case two, five-muon chains. The latter are therefore
more restricted in parameter space and more rate-suppressed.

The result is that the six-muon signature combines three advantages at
once: a large branching fraction over a broad part of the parameter space,
a very clean experimental final state, and enough internal structure to
reconstruct the cascade. This motivates the detailed study of the $6\mu$
channel in the rest of the paper.

\subsection{Six-muon final state}
\label{subsec:six_muon_final_state}

We now examine the structure of the six-muon final state in more detail.
The representative diagrams in \cref{fig:6mu-diags} realise the two
six-muon topologies contained in
\cref{eq:visible_muon_multiplicities}. Diagram (A) corresponds to an
asymmetric $(1+5)\mu$ topology: one $\mu'$ decay chain produces a single
muon together with invisible dark-sector particles, while the other chain
produces five visible muons through
\begin{equation}
\mu'\to \mu H_D,
\qquad
H_D\to V'V'\to 4\mu .
\end{equation}
Diagrams (B), (C), and (D) correspond to $(3+3)\mu$ topologies, in which
each $\mu'$ decay chain produces three visible muons. They differ by the
identity of the intermediate two-muon resonance: depending on the chain,
the two-muon system can originate from either $V'$ or $H_D$.

This topology decomposition is important because it determines the
reconstruction problem. In the symmetric $(3+3)\mu$ case, the event
contains two three-muon systems with comparable invariant masses,
corresponding to the two parent $\mu'$ states. This motivates a
reconstruction strategy based on pairing the six muons into two
three-muon candidates and looking for common two- and three-body resonance
structures. In the asymmetric $(1+5)\mu$ case, one side contains a
five-muon system while the other side contains a single muon and missing
transverse momentum from stable $V_D$ states. This topology requires a
separate reconstruction category rather than a forced symmetric assignment.

\begin{figure}[htb]
\centering
\includegraphics[trim={0cm 0cm 0 0cm 0cm},clip,width=0.99\textwidth]{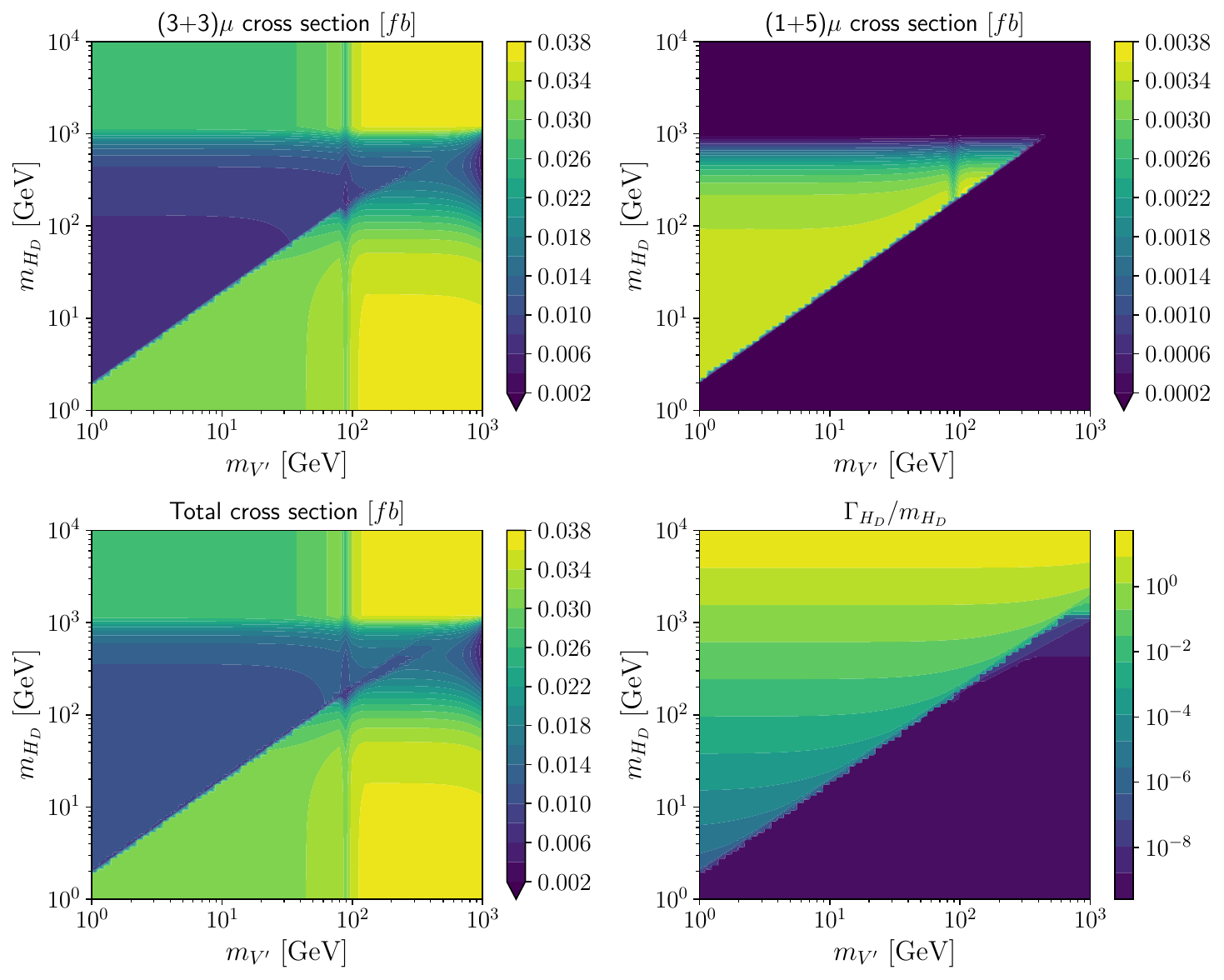}
\caption{\label{fig:xsec_6mu_plane}
Cross sections for representative six-muon final-state topologies and the
width-to-mass ratio of the dark scalar in the $(m_{V'},m_{H_D})$ plane.
The scan is performed for $m_{\mu'}=1200~\GeV$,
$m_{\mu_D}=1080~\GeV$, and $g_D=10m_{V'}/m_{\mu'}$.
The upper panels show the leading topology contributions, the lower-left
panel shows their combined contribution, and the lower-right panel shows
$\Gamma_{H_D}/m_{H_D}$. Regions with
$\Gamma_{H_D}/m_{H_D}>0.5$ are excluded from the analysis in order to avoid
loss of perturbative control and possible unitarity violation.}
\end{figure}

The corresponding six-muon cross sections are shown in
\cref{fig:xsec_6mu_plane}.
The figure is the rate-level counterpart of the diagrammatic classification
in \cref{fig:6mu-diags}. The upper panels show the leading six-muon topology
contributions, while the lower-left panel shows their sum.
The lower-right panel shows the ratio $\Gamma_{H_D}/m_{H_D}$. This quantity
is used as a perturbativity and unitarity-control diagnostic. When
$\Gamma_{H_D}/m_{H_D}$ becomes large, the dark scalar is no longer a
well-controlled perturbative state: the decay width is comparable to the
mass scale of the resonance, and the calculation enters a region where the
perturbative expansion and the usual particle interpretation become
unreliable. In the analysis below we therefore restrict ourselves to the
region
\begin{equation}
\frac{\Gamma_{H_D}}{m_{H_D}} < 0.5 ,
\label{eq:HD_width_cut}
\end{equation}
which removes the part of parameter space where perturbative control and
unitarity are potentially lost. The phenomenologically useful region is
therefore selected by two simultaneous requirements: an appreciable
six-muon rate and a controlled dark-scalar width.

The six-muon channel is therefore not chosen merely because it is exotic.
It is selected because it is both generic in the model and experimentally
powerful. The four-muon channel can be important, but it has a less
distinctive topology and faces larger SM and non-prompt backgrounds. The
eight- and ten-muon channels are cleaner, but their rates are more
restricted by the need for five-muon decay chains. The six-muon channel
sits in the optimal region between these alternatives: it is frequent enough
to probe a broad parameter space, clean enough to be nearly background-free
after suitable selections, and structured enough to allow reconstruction of
$V'$, $H_D$, and the parent $\mu'$.

The topology and rate considerations discussed above define the target of
the collider analysis. Before turning to detector simulation, we introduce
the benchmark scenarios used to test the reconstruction strategy across
representative regions of the $(m_{V'},m_{H_D})$ plane and for different
vector-like muon masses.

%%%%%%%%%%%%%%%%%%%%%%%%%%%%
%%%%%%%%%%%%%%%%%%%%%%%%%%%%%%

\subsection{Benchmark scenarios}
\label{subsec:benchmarks}

For the detector-level analysis we select representative benchmark scenarios
that illustrate the main kinematic regimes of the six-muon signature. The
benchmarks are not used to define the model; they are used to test the
performance of the reconstruction strategy in different regions of the
$(m_{V'},m_{H_D})$ plane.

The benchmarks are chosen to probe the complementary multimuon regime rather
than the large-splitting $2\mu+\met$ regime. We therefore focus on spectra
where $m_{\mu'}$ and $m_{\mu_D}$ are not widely separated, so that
$\mu'\to\mu_DV_D$ does not overwhelm the visible cascade modes
$\mu'\to\mu V'$ and $\mu'\to\mu H_D$. This choice is not an additional model
assumption, but a deliberate selection of the region where the multimuon
search provides information beyond the $2\mu+\met$ analysis of
Ref.~\cite{Belyaev:2025cgf}.

The inclusive production rate is controlled primarily by $m_{\mu'}$, while
the structure of the six-muon final state is controlled by the intermediate
dark-sector masses $m_{V'}$ and $m_{H_D}$. We therefore choose benchmark
points that span qualitatively different kinematic configurations: light
intermediate states, where the muons from $V'$ or $H_D$ can be collimated;
heavier intermediate states, where the muons are more separated; and regions
where the asymmetric and symmetric six-muon topologies contribute with
different weights. For each benchmark point we evaluate the physical branching
ratios of $\mu'$, $V'$ and $H_D$ with \texttt{CalcHEP}. The analytic
probabilities $P_1$, $P_3$ and $P_5$ introduced above are used only to
classify the possible muon multiplicities and the corresponding decay-chain
topologies.

The benchmark points are summarised in \cref{tab:benchmarks}. Besides the
input masses and $g_D$, we list the width-to-mass ratios of $\mu'$, $V'$, and
$H_D$, the total production cross section $\sigma(pp\to\mu'^{+}\mu'^{-})$,
the six-muon branching fraction, and the resulting parton-level six-muon
rate. These quantities specify both the validity of the resonance description
and the number of signal events available before detector acceptance and
reconstruction effects are applied.

The row labelled ``Contribution from $6\mu$ topologies $(3+3)+(1+5)$''
decomposes the total six-muon branching fraction into the symmetric $(3+3)\mu$
and asymmetric $(1+5)\mu$ cascade topologies as
\[
{\cal B}_{6\mu}={\cal B}_{(3+3)\mu}+{\cal B}_{(1+5)\mu}.
\]
This
distinction is important because the two topologies lead to different event
structures: the symmetric $(3+3)\mu$ case contains two three-muon systems
associated with the two parent $\mu'$ states, while the asymmetric $(1+5)\mu$
case contains one visible five-muon system and one branch with invisible
dark-sector particles. Thus BP1 and BP2 contain both topology classes, whereas
BP3--BP5 are dominated by the symmetric $(3+3)\mu$ topology.

\begin{table}[htb]
\centering
\setlength{\tabcolsep}{4.5pt}
\renewcommand{\arraystretch}{1.15}
\begin{tabular}{c|ccccc}
\toprule
 & BP1 & BP2 & BP3 & BP4 & BP5 \\
\midrule
$m_{\mu'}$ [GeV]    & 1200 & 1200 & 1200 & 1200 & 1200 \\
$m_{\mu_D}$ [GeV]   & 1080 & 1080 & 1080 & 1080 & 1080 \\
$m_{V'}$ [GeV]      & 1 & 10 & 100 & 100 & 300 \\
$m_{H_D}$ [GeV]     & 2.2 & 22 & 30 & 100 & 500 \\
$g_D$               & 0.00833 & 0.0833 & 0.833 & 0.833 & 2.5 \\
\midrule
$\Gamma_{\mu'}/m_{\mu'}$
 & $8.30\times10^{-2}$
 & $8.29\times10^{-2}$
 & $7.78\times10^{-2}$
 & $7.74\times10^{-2}$
 & $6.43\times10^{-2}$ \\
$\Gamma_{V'}/m_{V'}$
 & $8.00\times10^{-9}$
 & $8.27\times10^{-7}$
 & $8.27\times10^{-5}$
 & $8.28\times10^{-5}$
 & $7.45\times10^{-4}$ \\
$\Gamma_{H_D}/m_{H_D}$
 & $7.24\times10^{-7}$
 & $7.23\times10^{-5}$
 & $2.85\times10^{-10}$
 & $3.53\times10^{-10}$
 & $2.19\times10^{-9}$ \\
\midrule
$\sigma_{\mu'\mu'}$ [fb]   & 0.0320 & 0.0320 & 0.0320 & 0.0320 & 0.0320 \\
${\cal B}_{6\mu}$
 & 0.311 & 0.312 & 0.950 & 0.780 & 0.416 \\
 Contribution from  $6\mu$ topologies
$(3+3)+(1+5)$
 & $ 0.216+0.095$ & $0.216+0.096$ &  $0.950+0$
 &  $0.780+0$ &   $0.416+0$ \\
$\sigma_{6\mu}$ [fb]
 & 0.0100 & 0.0100 & 0.0304 & 0.0249 & 0.0133 \\
\bottomrule
\end{tabular}
\caption{\label{tab:benchmarks}
Benchmark scenarios used for the detector-level six-muon analysis.
Here $\sigma_{\mu'\mu'}\equiv\sigma(pp\to\mu'^{+}\mu'^{-})$ and
$\sigma_{6\mu}=\sigma_{\mu'\mu'}{\cal B}_{6\mu}$ before detector acceptance
and reconstruction efficiencies. The width-to-mass ratios quantify
the validity of the resonance description and show, in particular, that all
benchmarks satisfy $\Gamma_{H_D}/m_{H_D}<0.5$.}
\end{table}
The benchmark points therefore connect the kinematic structure of the
$(m_{V'},m_{H_D})$ plane to the detector-level analysis below. They probe
different mediator-mass configurations, different degrees of muon collimation,
and different mixtures of the symmetric $(3+3)\mu$ and asymmetric $(1+5)\mu$
six-muon topologies.

In the following section we simulate these benchmark scenarios at detector
level, apply the multimuon event selection, and develop the reconstruction
strategy used to identify the intermediate $V'$, $H_D$, and parent $\mu'$
states.

%%%%%%%%%%%%%%%%%%%%%%%%%%%%%%%%%%%%%%%%%%%%%%%%%%%%%%%%%%%%%%%%%%
\section{Collider analysis}

\label{sec:collider_analysis}
%%%%%%%%%%%%%%%%%%%%%%%%%%%%%%%%%%%%%%%%%%%%%%%%%%%%%%%%%%%%%%%%%%%%%%%%%%%%%%%

We now develop a detector-level analysis of the six-muon signal. The analysis
has two complementary aims. First, we determine the constraints already
implied by existing LHC searches through a \texttt{CheckMATE} recast.
Secondly, we construct a dedicated topology-based search which exploits the
two- and three-body resonance structure of the signal and determine its
expected reach with present and HL-LHC luminosities.

%%%%%%%%%%%%%%%%%%%%%%%%%%%%%%%%%%%%%%%%%%%%%%%%%%%%%%%%%%%%%%%%%%%%%%%%%%%%%%%
\subsection{Event generation and detector simulation}
\label{subsec:simulation}
%%%%%%%%%%%%%%%%%%%%%%%%%%%%%%%%%%%%%%%%%%%%%%%%%%%%%%%%%%%%%%%%%%%%%%%%%%%%%%%

The event-generation and detector-simulation setup used
\texttt{CalcHEP}~4.1, \mg~3.6.3, \pythia~8.317,
\texttt{Delphes}~3.5.1 and \texttt{CheckMATE}~2.0.41.
Signal events were generated at LO with the MPVDM implementation in
\texttt{CalcHEP}. The benchmark samples introduced in
\cref{subsec:benchmarks} were used to study the event kinematics and validate
the reconstruction procedure, while parameter grids in
$(m_{V'},m_{H_D})$ were generated for the limit calculations. The
$t\bar t$ background sample was also generated with \texttt{CalcHEP}.
The \texttt{CalcHEP} implementation of the MPVDM model is available from
HEPModelDB at \url{https://hepmdb.soton.ac.uk/hepmdb:1225.0357}.
The remaining Standard Model background samples were generated with
\mg~\cite{Alwall:2014hca}.

All generated events were stored in the Les Houches Event (LHE) format and passed
to \pythia~\cite{Bierlich:2022pfr} for parton showering, hadronisation, and
unstable-particle decays. Detector effects were simulated with
\texttt{Delphes}~\cite{deFavereau:2013fsa}.
Two detector-simulation chains were used for distinct purposes. Existing
Run-2 constraints were evaluated with
\texttt{CheckMATE}~\cite{Drees:2013wra,Kim:2015wza,Dercks:2016npn}, which automates the
LHE--\pythia--\texttt{Delphes} chain and applies the implemented
experimental analyses. For the CMS recast discussed in
\cref{subsec:current_limits}, we corrected the CMS \texttt{Delphes} card
used by \texttt{CheckMATE}, whose original muon momentum-resolution
parameterisation produced an incorrect resolution in the kinematic range
relevant here. The dedicated reconstruction study and the HL-LHC projection
were instead performed with an ATLAS-based \texttt{Delphes} setup. The
ATLAS and CMS muon systems have closely comparable performance for the
energetic, central muons that dominate this analysis, while the dedicated
study depends mainly on the reconstruction of correlated dimuon and trimuon
resonances rather than on detector-specific hadronic observables.
Unless stated otherwise, reconstructed muons are required to satisfy
\begin{equation}
p_T^\mu>10~\GeV,
\qquad
|\eta_\mu|<2.5 .
\label{eq:muon_baseline}
\end{equation}
The signal preselection requires at least six reconstructed muons. No
requirement on jets, $b$ tags, or $\met$ is imposed at this stage. This keeps
the selection inclusive with respect to the fully visible $(3+3)\mu$
topologies and the $(1+5)\mu$ topologies containing invisible $V_D$ states.
%%%%%%%%%%%%%%%%%%%%%%%%%%%%%%%%%%%%%%%%%%%%%%%%%%%%%%%%%%%%%%%%%%%%%%%%%%%%%%%
\subsection{Standard Model backgrounds}
\label{subsec:backgrounds}
%%%%%%%%%%%%%%%%%%%%%%%%%%%%%%%%%%%%%%%%%%%%%%%%%%%%%%%%%%%%%%%%%%%%%%%%%%%%%%%

The irreducible background is continuum production of six prompt muons,
\begin{equation}
pp\to 3(\mu^+\mu^-).
\end{equation}
Additional backgrounds arise from processes containing prompt muons from
electroweak decays together with non-prompt muons from heavy-flavour hadron
decays. We consider
\begin{align}
pp&\to ZZ\,W^+W^-
 \to 6\mu+2\nu, \label{eq:bg_zzww}\\
pp&\to t\bar t
 \to 2\mu+2\nu+2b, \label{eq:bg_tt}\\
pp&\to t\bar t Z
 \to 4\mu+2\nu+2b, \label{eq:bg_ttz}\\
pp&\to t\bar t t\bar t
 \to 4\mu+4\nu+4b. \label{eq:bg_4t}
\end{align}
Parton showering, hadronisation, and heavy-flavour decays were
performed with \pythia. For the \mg{} background samples we used the
\texttt{NNPDF23\_nlo\_as\_0119} PDF set~\cite{Ball:2012cx} through
\lhapdf, corresponding to LHAPDF identifier 230000.

The probability for heavy-flavour decays to provide the additional muons
required by the six-muon selection is small. Direct generation would therefore
lead to poor Monte Carlo statistics. We therefore use a repeated-decay filter
in which the weakly decaying bottom hadrons in each showered and hadronised
event are decayed many times with the physical \pythia{} decay tables. The
code keeps the \texttt{moreDecays} mode as a runtime option. The production
samples use \texttt{moreDecays(false)}, which is the default treatment in the
\pythia{} version used here; the cross-check with \texttt{moreDecays(true)} is
described in \cref{app:background_normalisation}.

The $t\bar t Z$ and four-top samples already contain four prompt muons, so
the heavy-flavour filter requires at least two additional muons from
bottom-hadron decay chains. The $t\bar t$ sample contains two prompt muons and
therefore requires at least four additional heavy-flavour muons. In the
filtering step an additional muon is accepted only if it descends from one of
the weakly decaying bottom-hadron chains identified in the event record,
satisfies $p_T^\mu>10~\GeV$ and $|\eta_\mu|<2.5$, and passes the
charged-particle isolation requirement
\begin{equation}
\frac{\sum_{\Delta R<0.5}p_T^{\rm charged}}{p_T^\mu}<0.5,
\qquad p_T^{\rm charged}>0.5~\GeV .
\label{eq:hf_muon_isolation}
\end{equation}
All charged final-state particles other than the candidate muon are included
in the isolation sum. This generator-level requirement is looser than the
final \texttt{Delphes} muon isolation (0.25),  which is applied to particle-flow
objects in the detector simulation. It is used only to enrich the rare
heavy-flavour samples before the final reconstructed-muon selection.

The heavy-flavour filtering efficiencies and the corresponding filtered cross
sections are summarised in \cref{tab:background_generation}; the detailed
normalisation formulae are given in \cref{app:background_normalisation}. The
$t\bar t$ filter cross section is already below $10^{-7}~{\rm fb}$ before the
final detector-level six-muon selection, and this background is therefore not
propagated through the subsequent reconstruction. The surviving heavy-flavour
events from the $t\bar t Z$ and four-top samples, together with the continuum
$6\mu$ background, are further reduced by the topological reconstruction and
mass-window requirements, since their muon combinations do not reproduce the
correlated resonance structure of the signal.

The dedicated code developed to generate and filter physical heavy-flavour
decays, together with the topology-reconstruction and event-selection code
used in this analysis, is provided as supplementary material. The
heavy-flavour package includes the source file and Makefile required for
compilation within \texttt{Delphes}.

\begin{table}[htbp]
\centering
\renewcommand{\arraystretch}{1.3}
\begin{tabular}{l c c c c}
\toprule
\textbf{Generated background process}
&
\textbf{$\sigma_{\rm ME}$ [fb]}
&
\textbf{Required $N_{\mu}^{\rm HF}$}
&
\textbf{$\epsilon_{\rm HF}$}
&
\textbf{$\sigma_{\rm ME}\epsilon_{\rm HF}$ [fb]}
\\
\midrule
$pp\to t\bar t\to 2\mu+2\nu+2b$
&
$2.74\times10^{3}$
&
$\geq4$
&
$3.0\times10^{-11}$
&
$8.21\times 10^{-8}$
\\
$pp\to t\bar t Z\to 4\mu+2\nu+2b$
&
$0.248$
&
$\geq2$
&
$5.49\times10^{-5}$
&
$1.36\times10^{-5}$
\\
$pp\to t\bar t t\bar t\to 4\mu+4\nu+4b$
&
$1.41\times10^{-3}$
&
$\geq2$
&
$2.38\times10^{-4}$
&
$3.36\times10^{-7}$
\\
$pp\to ZZ\,W^+W^-\to 6\mu+2\nu$
&
$5.36\times10^{-6}$
&
$0$
&
$--$
&
$5.36\times10^{-6}$
\\
$pp\to 6\mu$
&
$8.61\times10^{-5}$
&
$0$
&
$--$
&
$8.61\times10^{-5}$
\\
\bottomrule
\end{tabular}
\caption{Standard Model background samples used in the $6\mu$ analysis.
The quantity $\sigma_{\rm ME}$ is the cross section of the explicitly
generated matrix-element final state shown in the first column, before
heavy-flavour decays. The third column gives the minimum number
$N_{\mu}^{\rm HF}$ of additional muons required from heavy-flavour decay
chains. The $t\bar t$ sample contains two prompt muons and therefore requires
at least four additional heavy-flavour muons, whereas the $t\bar t Z$ and
four-top samples contain four prompt muons and require at least two additional
heavy-flavour muons. The corresponding probability is denoted by
$\epsilon_{\rm HF}$, and the final column gives the cross section after this
generator-level requirement,
$\sigma_{\rm filter}=\sigma_{\rm ME}\epsilon_{\rm HF}$. The
$ZZ\,W^+W^-\to6\mu+2\nu$ and continuum $6\mu$ samples already contain six
prompt muons and require no additional heavy-flavour muons.}
\label{tab:background_generation}
\end{table}

%%%%%%%%%%%%%%%%%%%%%%%%%%%%%%%%%%%%%%%%%%%%%%%%%%%%%%%%%%%%%%%%%%%%%%%%%%%%%%%
\subsection{Signal kinematics and resonance structure}
\label{subsec:signal_kinematics}
%%%%%%%%%%%%%%%%%%%%%%%%%%%%%%%%%%%%%%%%%%%%%%%%%%%%%%%%%%%%%%%%%%%%%%%%%%%%%%%

\begin{figure}[htbp]
\centering
\includegraphics[width=0.5\textwidth]{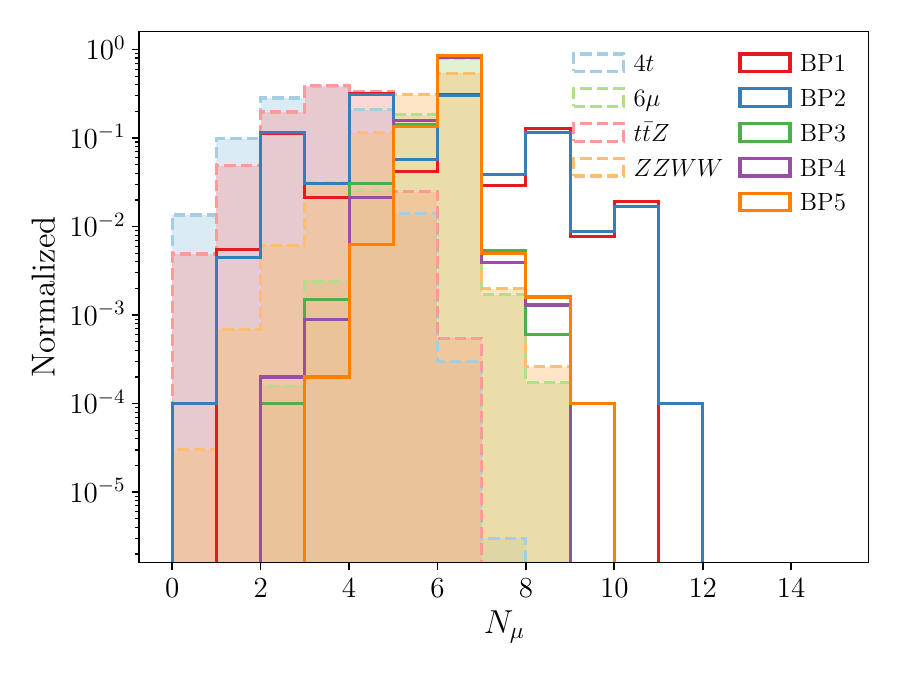}%
\includegraphics[width=0.5\textwidth]{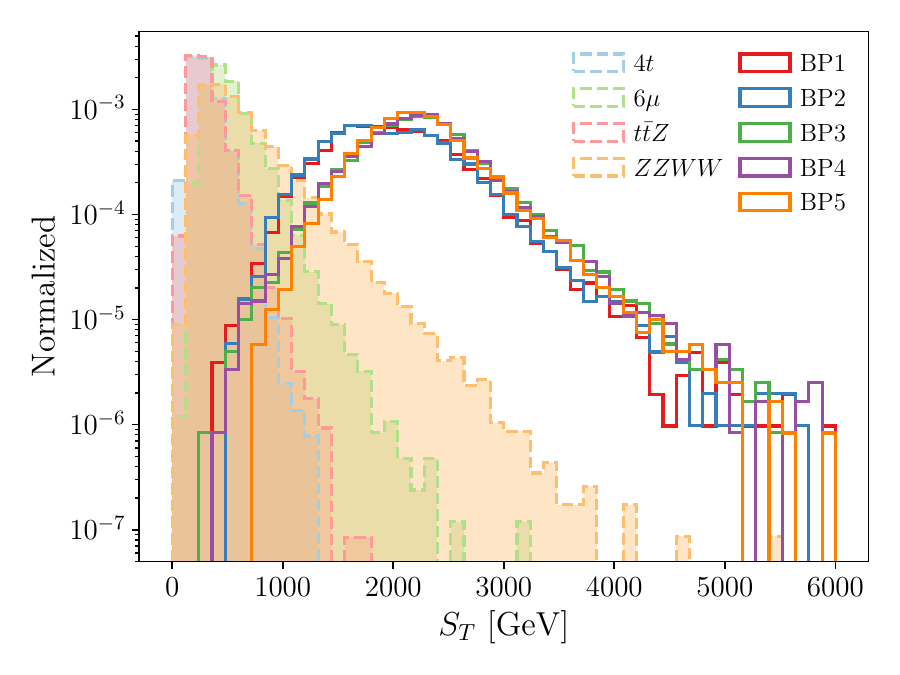}\vskip 0.7cm
\includegraphics[width=0.5\textwidth]{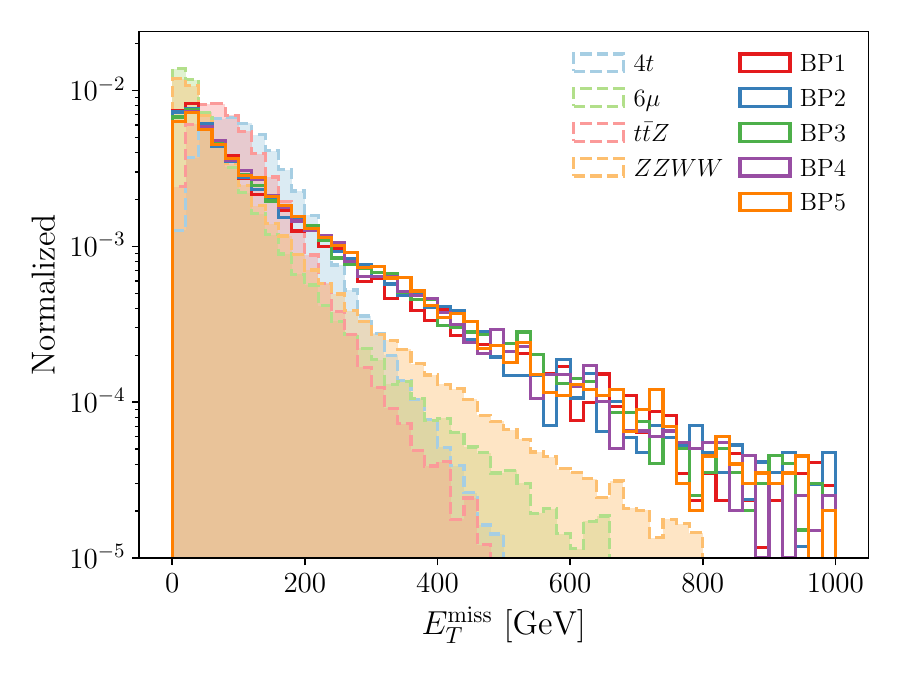}%
\includegraphics[width=0.5\textwidth]{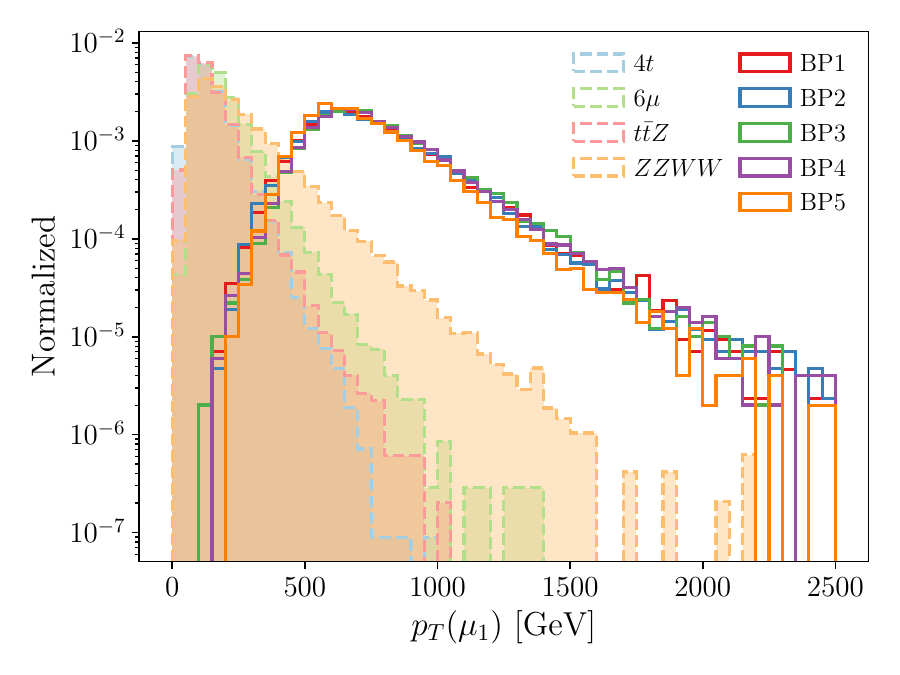}
\caption{\label{fig:global_kinematics}
Normalised kinematic distributions for the five signal benchmarks and the
dominant SM backgrounds. Top left: reconstructed muon multiplicity. Top
right: $S_T^\mu$. Bottom left: $\met$. Bottom right: transverse momentum of
the leading muon. The requirement $p_T^\mu>10~\GeV$ and
$|\eta_\mu|<2.5$ is applied; the multiplicity distribution is shown before
the requirement $N_\mu\geq6$. }
\end{figure}

\begin{figure}[htbp]
\centering
\includegraphics[width=0.48\textwidth]{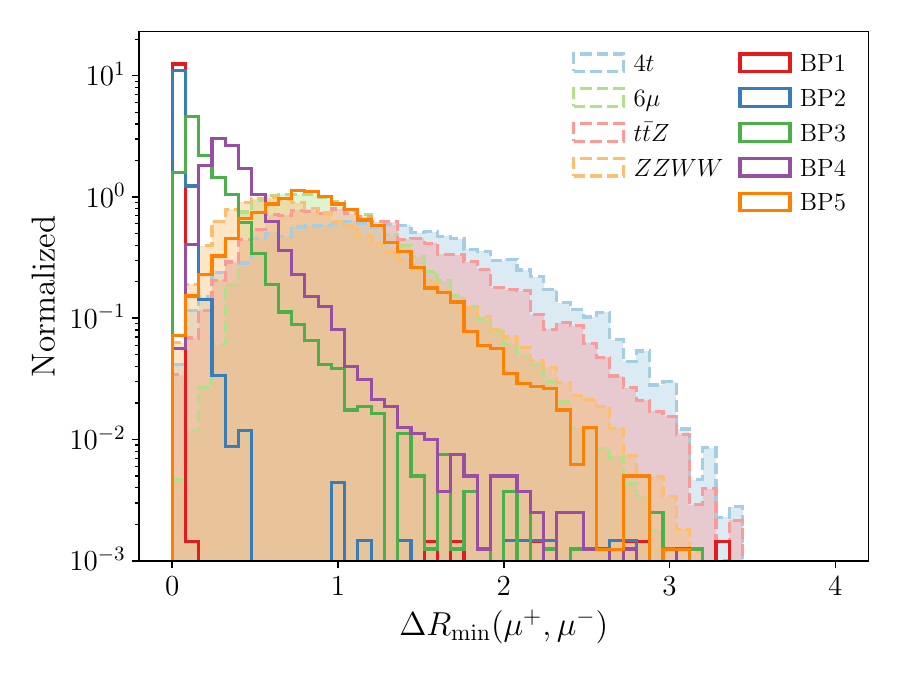}
\includegraphics[width=0.48\textwidth]{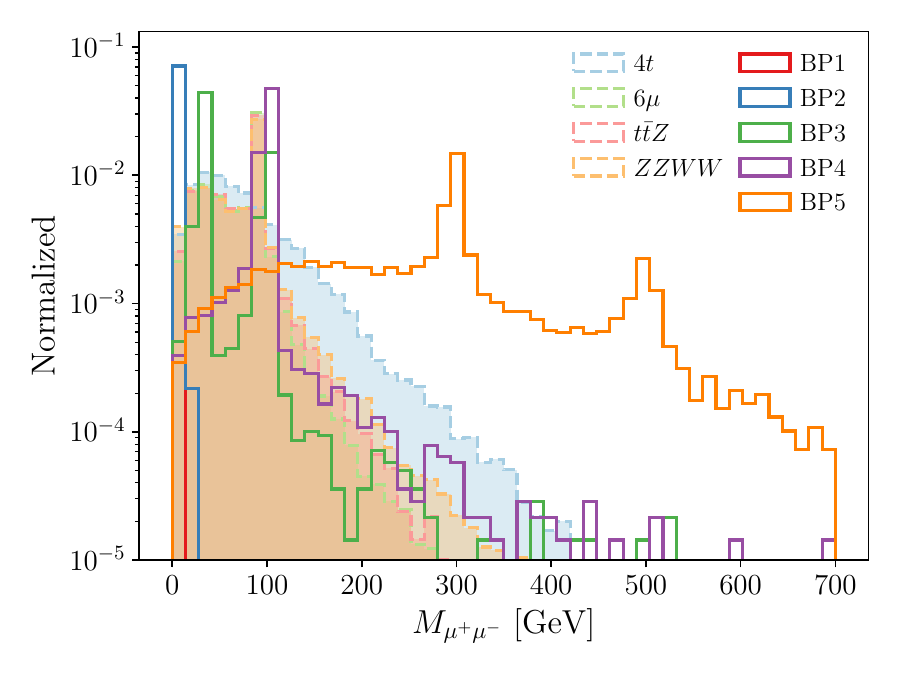}
\caption{\label{fig:substructure}
Resonant substructure of the signal cascades. Left: minimum angular
separation between opposite-sign muons. Right: invariant mass of the closest
opposite-sign muon pair. The light-mediator benchmarks give collimated
dimuons, while the reconstructed dimuon masses retain the intermediate
resonance information. }
\end{figure}

The main detector-level features of the signal are illustrated in
\cref{fig:global_kinematics,fig:substructure} for the five benchmark points
of \cref{tab:benchmarks} and the backgrounds described above. The
distributions are normalised to unit area in order to compare their shapes.
The muon multiplicity in \cref{fig:global_kinematics} is shown before
applying the six-muon requirement; the remaining distributions use the
baseline muon selection in \cref{eq:muon_baseline}.

Figure~\ref{fig:global_kinematics} shows the inclusive event-level
kinematics. The signal benchmarks populate the high-muon-multiplicity tail,
whereas the SM backgrounds fall rapidly beyond four reconstructed muons.
This behaviour reflects the cascade structure of the signal: six or more
muons are produced directly in the hard decay chain, while the SM
backgrounds require either rare prompt multiboson configurations or
additional isolated muons from heavy-flavour decays. The scalar sum of the
muon transverse momenta,
\begin{equation}
S_T^\mu=\sum_{\mu}p_T^\mu ,
\label{eq:stmu}
\end{equation}
is also substantially harder for the signal because it inherits the
$O(\TeV)$ mass scale of the pair-produced $\mu'$ states. The same feature is
visible in the leading-muon transverse-momentum distribution: the leading
muon is often produced directly in the two-body decay of a heavy $\mu'$, and
therefore carries a large fraction of the heavy-lepton mass scale.

By contrast,  the missing transverse momentum, $\met$, is not a universal discriminator for this signal. It is
sizeable for the asymmetric $(1+5)\mu$ topology, where one branch contains
invisible $V_D$ states, but it is much smaller for the fully visible
$(3+3)\mu$ topology. A common $\met$ requirement would therefore remove a
significant part of the signal parameter space. We consequently keep the
preselection inclusive in $\met$ and instead exploit the resonance
substructure of the muon system.

The resonance-level information is shown in \cref{fig:substructure}. The
minimum angular separation between opposite-sign muons is particularly
sensitive to the mass of the intermediate state. For the light-mediator
benchmarks, especially BP1 with $m_{V'}=1~\GeV$, the muons from
$V'\to\mu^+\mu^-$ are highly collimated and form a lepton-jet-like
configuration with very small $\Delta R$. As $m_{V'}$ increases, the
opposite-sign muons become more separated, as illustrated by the shift of the
BP3--BP5 distributions to larger $\Delta R$.

The invariant mass of the closest opposite-sign muon pair retains the
resonance information from the cascade. Signal events show structures
associated with the intermediate $V'$ or $H_D$ states, while the continuum
and heavy-flavour backgrounds are much smoother and concentrated at lower
masses. These features show that the signal should not be treated only as a
high-multiplicity counting experiment. The repeated dimuon and higher-body
mass structures provide the information needed to resolve the cascade
topology and motivate the reconstruction strategy developed in the next
subsection.

%%%%%%%%%%%%%%%%%%%%%%%%%%%%%%%%%%%%%%%%%%%%%%%%%%%%%%%%%%%%%%%%%%%%%%%%%%%%%%%

\subsection{Topology-based event reconstruction}
\label{subsec:topology_reconstruction}
%%%%%%%%%%%%%%%%%%%%%%%%%%%%%%%%%%%%%%%%%%%%%%%%%%%%%%%%%%%%%%%%%%%%%%%%%%%%%%%

The reconstruction is organised into two exclusive signal categories
corresponding to the two six-muon topologies identified in
\cref{subsec:six_muon_final_state}. Category~1 targets the fully visible
symmetric $(3+3)\mu$ topology, while Category~2 recovers the asymmetric
$(1+5)\mu$ topology.

The topology reconstruction is applied after detector simulation. The first
analysis requirement is therefore a reconstructed six-muon preselection,
\[
N_\mu^{\rm reco}\geq 6,
\]
where the muons satisfy the baseline requirements in
\cref{eq:muon_baseline} and the final Delphes isolation. This is distinct
from the generator-level heavy-flavour filtering described in
\cref{subsec:backgrounds}, where additional heavy-flavour muons are selected
at truth level after showering and hadronisation using a looser charged
isolation requirement.

For each topology, we enumerate the allowed assignments of reconstructed
muons to the visible decay products. Each assignment is scored using a sum
of $\chi^2$ compatibility terms, with each term testing one resonance
condition. The best combinatorial assignment is selected by minimising the
corresponding total $\chi^2$, after which the event selection is applied to
the parent-mass candidate reconstructed from that assignment.

The explicit definitions of the individual $\chi^2$ terms are given in
Appendix~\ref{app:chi2_details},
Eqs.~\eqref{eq:chi2_two_masses_appendix} and
\eqref{eq:chi2_target_appendix}. The logarithmic mass differences make the
comparison depend on relative rather than absolute mass resolution and are
therefore appropriate for scans extending from GeV-scale mediators to
TeV-scale parent particles.

\paragraph{Category 1: symmetric $(3+3)\mu$ reconstruction.}
The six muons are partitioned into two charge-compatible trimuon systems.
Within each trimuon system, an opposite-sign pair is assigned to the
intermediate $V'$ or $H_D$ resonance. For each allowed assignment, we form
two dimuon masses, $m_{2\mu}^{(1,2)}$, and two trimuon masses,
$m_{3\mu}^{(1,2)}$. The preferred assignment minimises
\begin{equation}
\chi^2_{\rm C1}
=
\chi^2_{2\mu}\!\left(m_{2\mu}^{(1)},m_{2\mu}^{(2)}\right)
+
\chi^2_{3\mu}\!\left(m_{3\mu}^{(1)},m_{3\mu}^{(2)}\right),
\label{eq:chi2_cat1}
\end{equation}
where the two terms test, respectively, the compatibility of the two
intermediate dimuon masses and of the two reconstructed parent masses. The
assignment therefore does not require the numerical values of $m_{V'}$,
$m_{H_D}$, or $m_{\mu'}$ to be specified in advance.

\begin{figure}[htbp]
\centering
\includegraphics[width=0.95\textwidth]{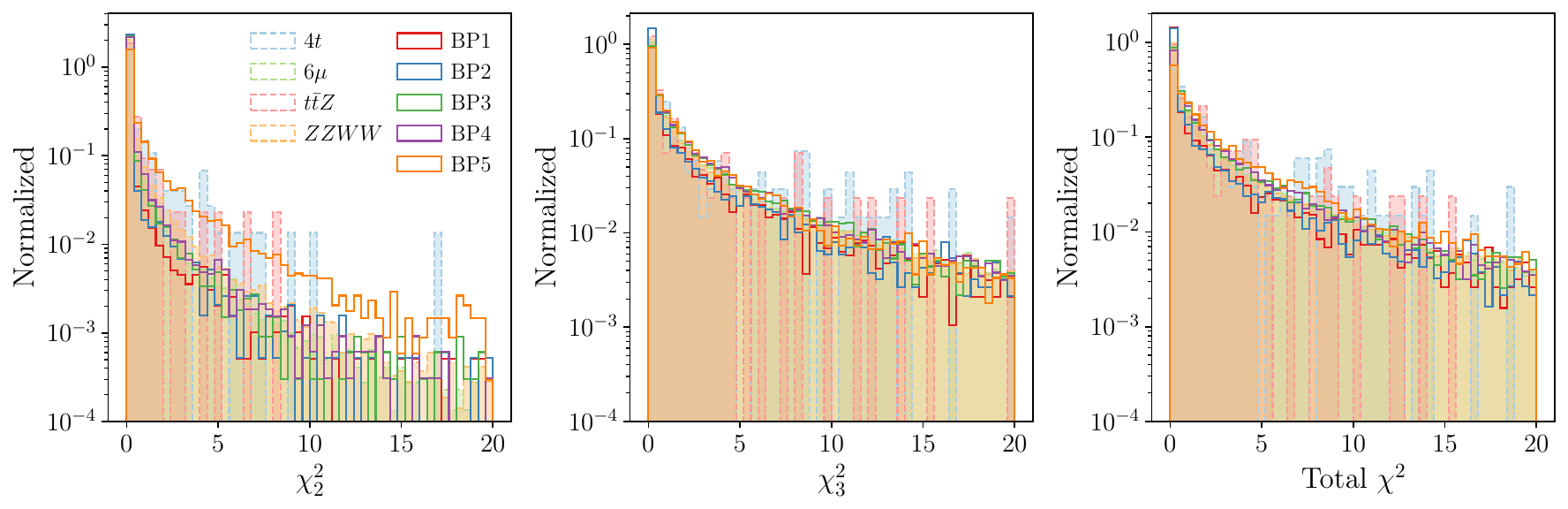}
\caption{\label{fig:cat1_chi2}
Category-1 reconstruction variables for the benchmark signals and SM
backgrounds. From left to right: the dimuon compatibility term, the trimuon
compatibility term, and their sum.}
\end{figure}

\begin{figure}[htbp]
\centering
\includegraphics[width=0.95\textwidth]{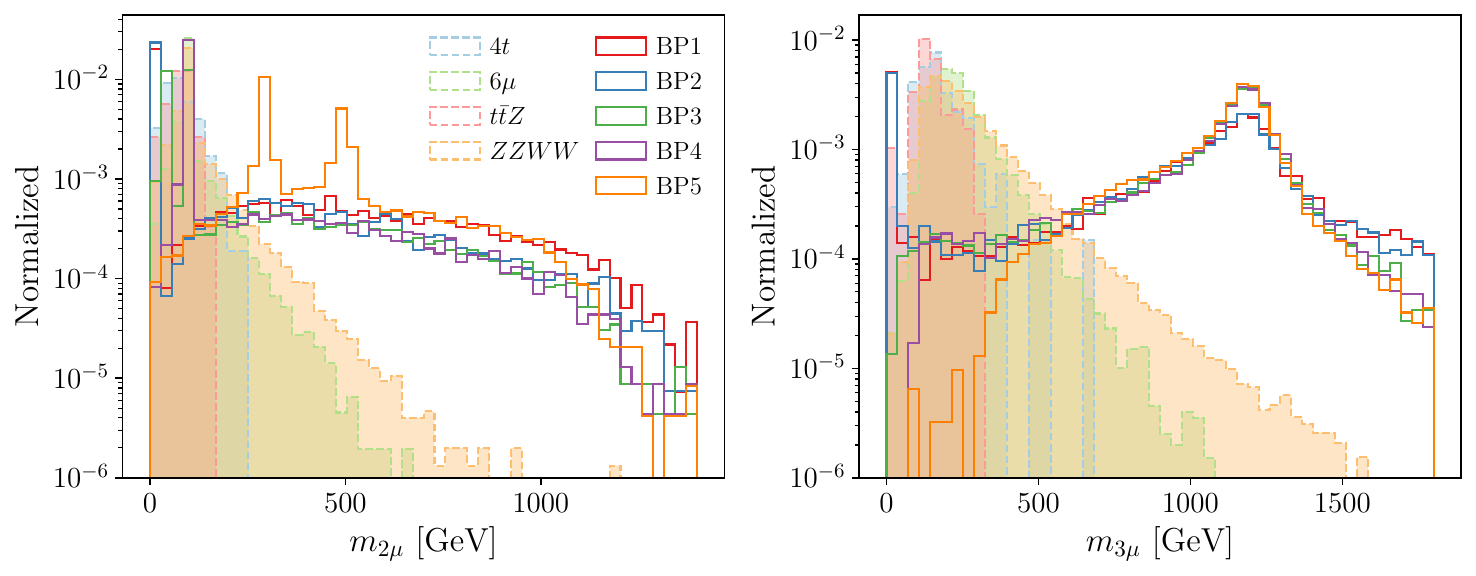}
\caption{\label{fig:cat1_mass}
Invariant masses selected by the Category-1 assignment. Left: reconstructed
dimuon resonance mass. Right: reconstructed trimuon parent mass. The
trimuon peak reconstructs $m_{\mu'}$, while the dimuon structure probes
$m_{V'}$ or $m_{H_D}$.}
\end{figure}

The Category-1 $\chi^2$ compatibility variables are shown in
\cref{fig:cat1_chi2}. Their main role is to select the most consistent
combinatorial assignment rather than to act as stand-alone
signal-background discriminants. Once this assignment has been selected,
the reconstructed dimuon and trimuon masses are shown in
\cref{fig:cat1_mass}. In particular, the right panel displays a clear
$m_{3\mu}$ peak around the parent mass $m_{\mu'}$ and motivates the
parent-mass selection below.

After selecting the assignment that minimises $\chi^2_{\rm C1}$, the event
is accepted into Category~1 only if both reconstructed trimuon masses lie
within a window around the target heavy-lepton mass:
\begin{equation}
\left|m_{3\mu}^{(i)}-m_{\mu'}^{\rm target}\right|
<
0.2\,m_{\mu'}^{\rm target},
\qquad i=1,2 .
\label{eq:cat1_mass_window}
\end{equation}

The choice of a $\pm20\%$ mass window is indicative rather than fully
optimised. As can be seen from the right panel of
\cref{fig:cat1_mass}, it retains approximately $80\%$ of the reconstructed
signal under the $m_{3\mu}$ peak. This is sufficient for the present
analysis. The window could nevertheless be optimised and enlarged further, particularly because the residual SM background is
already extremely small.
This optimisation will be relevant to a concrete experimental analysis using the detector-specific mass resolution and
the observed shape of the reconstructed peak.

Events failing the double-trimuon mass-window requirement are passed to the
asymmetric Category~2 reconstruction rather than being discarded. No
separate numerical threshold is imposed on $\chi^2_{\rm C1}$ itself or on
the reconstructed dimuon masses. The rejection power of Category~1 therefore
comes from requiring both reconstructed parent candidates to lie within the
target $\mu'$ mass window after the best symmetric assignment has been
selected.

For the benchmark samples, $m_{\mu'}^{\rm target}$ is set equal to the
generated value of $m_{\mu'}$. In an experimental search, the same procedure
would be implemented as a scan over the reconstructed parent-mass spectrum.

\paragraph{Category 2: asymmetric $(1+5)\mu$ reconstruction.}
Only events that fail the Category~1 selection are tested against the
asymmetric topology. Five muons are combined into a visible parent candidate,
while the sixth muon is assigned to the branch containing invisible $V_D$
states. The five-muon system contains two opposite-sign dimuon resonance
candidates. For each allowed assignment, we form the two dimuon masses,
$m_{2\mu}^{(1,2)}$, and the five-muon mass, $m_{5\mu}$. The preferred
assignment minimises
\begin{equation}
\chi^2_{\rm C2}
=
\chi^2_{2\mu}\!\left(m_{2\mu}^{(1)},m_{2\mu}^{(2)}\right)
+
\chi^2_{5\mu}\!\left(m_{5\mu},m_{\mu'}^{\rm target}\right),
\label{eq:chi2_cat2}
\end{equation}
where the first term tests the compatibility of the two dimuon resonances
inside the visible five-muon branch, while the second compares the
reconstructed five-muon mass with the target $\mu'$ mass. In an experimental
application, the target mass can be scanned or inferred from a Category-1
excess. The explicit uncertainty propagation and the definitions of the
individual $\chi^2$ terms are given in
Appendix~\ref{app:chi2_details}.

After selecting the assignment that minimises $\chi^2_{\rm C2}$, the event
is accepted into Category~2 only if the reconstructed five-muon parent mass
lies within the same $\pm20\%$ window around the target heavy-lepton mass,
\begin{equation}
\left|m_{5\mu}-m_{\mu'}^{\rm target}\right|
<
0.2\,m_{\mu'}^{\rm target}.
\label{eq:cat2_mass_window}
\end{equation}
Events failing this five-muon mass-window requirement have already failed
the Category~1 double-trimuon requirement and are therefore rejected from the
topology-selected sample. Since Category~2 is applied only after
Category~1 has failed, the two categories are exclusive by construction.
No separate numerical threshold is imposed on $\chi^2_{\rm C2}$ itself or
on the reconstructed dimuon masses. The full selection logic is therefore
\[
N_\mu^{\rm reco}\geq6
\quad\longrightarrow\quad
{\rm Category~1}
\quad\longrightarrow\quad
{\rm Category~2~only~if~Category~1~fails}.
\]
An event is retained if it passes either the Category~1 parent-mass window in
\cref{eq:cat1_mass_window} or, after failing Category~1, the Category~2
parent-mass window in \cref{eq:cat2_mass_window}. Events failing both
parent-mass requirements are rejected.

\begin{figure}[htbp]
\centering
\includegraphics[width=0.95\textwidth]{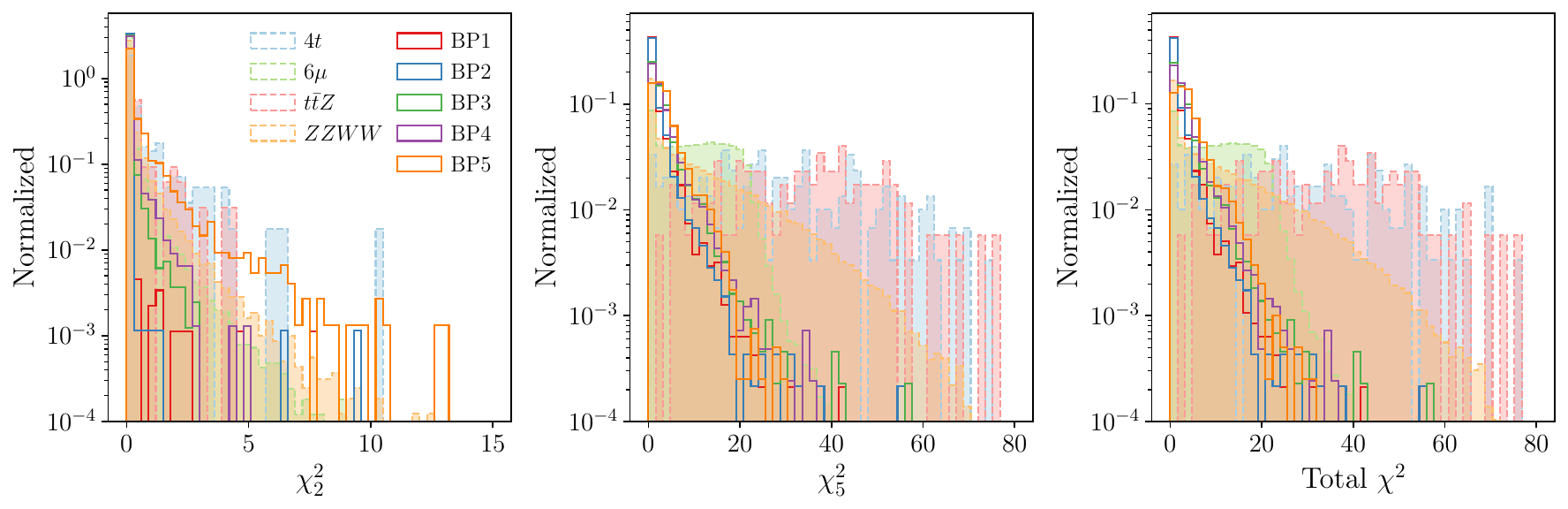}
\caption{\label{fig:cat2_chi2}
Category-2 reconstruction variables for the benchmark signals and SM
backgrounds. From left to right: the dimuon compatibility term, the
five-muon parent term, and their sum.}
\end{figure}

\begin{figure}[htbp]
\centering
\includegraphics[width=0.95\textwidth]{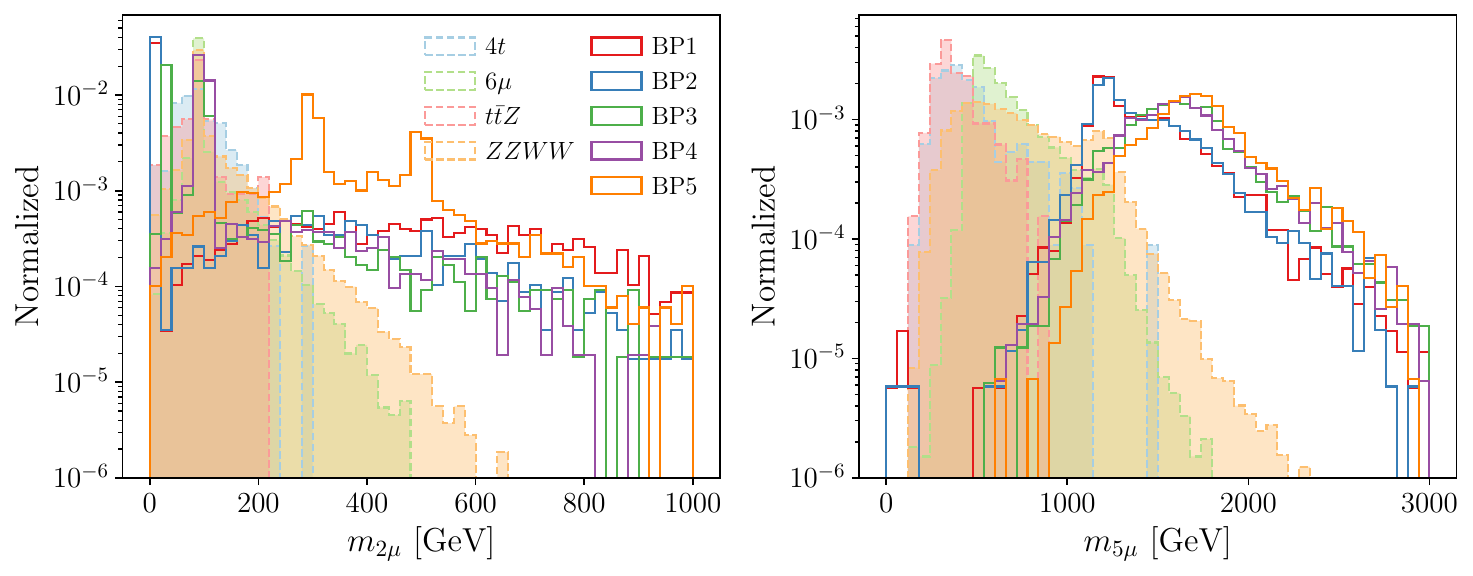}
\caption{\label{fig:cat2_mass}
Invariant masses selected by the Category-2 assignment. Left: reconstructed
dimuon resonance mass. Right: reconstructed five-muon parent mass.}
\end{figure}

The corresponding Category-2 distributions are shown in
\cref{fig:cat2_chi2,fig:cat2_mass}. The five-muon invariant mass reconstructs
the visible $\mu'$ branch, while the dimuon masses retain the nested
$H_D\to V'V'$ resonance structure. Category~2 is less selective than
Category~1 because only one heavy parent is reconstructed and the other
branch contains missing momentum. It nevertheless preserves sensitivity to
the $(1+5)\mu$ topology, which would be discarded by a purely symmetric
analysis.

The benchmark selection efficiencies are shown in \cref{tab:cutflow_eff}.
The corresponding expected event yields at $3000~\mathrm{fb}^{-1}$ are
shown in \cref{tab:cutflow_yields}. For entries with zero selected Monte
Carlo events, the table gives the upper value obtained by replacing zero
with one selected event.

\begin{table}[htbp]
\centering
\setlength{\tabcolsep}{3.0pt}
\renewcommand{\arraystretch}{1.15}
\begin{tabular}{lccccccccc}
\toprule
Selection & BP1 & BP2 & BP3 & BP4 & BP5 &
$4t$ & $6\mu$ & $t\bar tZ$ & $ZZWW$\\
\midrule
$N_\mu^{\rm reco}\geq6$ & 0.906 & 0.863 & 0.823 & 0.832 & 0.862 &
$3.02\times10^{-4}$ & 0.789 & $5.46\times10^{-4}$ & 0.538\\
Category~1 & 0.354 & 0.323 & 0.538 & 0.553 & 0.606 &
$<1.0\times10^{-6}$ & $8.57\times10^{-5}$ & $<3.64\times10^{-6}$ &
$1.01\times10^{-3}$\\
Category~2 & 0.306 & 0.299 & 0.0690 & 0.0700 & 0.0381 &
$8.0\times10^{-6}$ & 0.0961 & $<3.64\times10^{-6}$ & 0.133\\
\bottomrule
\end{tabular}
\caption{\label{tab:cutflow_eff}
Selection efficiencies for the benchmark signals and SM backgrounds. The
first row requires at least six reconstructed Delphes muons satisfying the
baseline selection and final isolation requirements. This is distinct from
the generator-level heavy-flavour filtering used to enrich the $t\bar tZ$
and four-top samples. The Category~1 row gives the efficiency after the
symmetric reconstruction and double-trimuon parent-mass window. The
Category~2 row is evaluated only for events failing Category~1 and gives the
additional exclusive efficiency after the asymmetric reconstruction and
five-muon parent-mass window. The $t\bar t$ sample is not included because
its filtered cross section is already negligible before this stage.
The signal efficiencies are defined relative to events containing at
least six visible muons at truth level.}
\end{table}

\begin{table}[htbp]
\centering
\setlength{\tabcolsep}{3.0pt}
\renewcommand{\arraystretch}{1.15}
\begin{tabular}{lccccccccc}
\toprule
Selection & BP1 & BP2 & BP3 & BP4 & BP5 &
$4t$ & $6\mu$ & $t\bar tZ$ & $ZZWW$\\
\midrule
Initial & 47.5 & 48.0 & 91.2 & 74.7 & 39.9 &
$1.01\times10^{-3}$ & 0.258 & $4.08\times10^{-2}$ &
$1.61\times10^{-2}$\\
$N_\mu^{\rm reco}\geq6$ & 43.0 & 41.4 & 75.0 & 62.1 & 34.4 &
$3.04\times10^{-7}$ & 0.204 & $2.23\times10^{-5}$ &
$8.66\times10^{-3}$\\
Category~1 & 16.8 & 15.5 & 49.0 & 41.3 & 24.2 &
$<1.01\times10^{-9}$ & $2.21\times10^{-5}$ & $<1.49\times10^{-7}$ &
$1.62\times10^{-5}$\\
Category~2 & 14.5 & 14.4 & 6.29 & 5.23 & 1.52 &
$8.06\times10^{-9}$ & $2.48\times10^{-2}$ & $<1.49\times10^{-7}$ &
$2.13\times10^{-3}$\\
\bottomrule
\end{tabular}
\caption{\label{tab:cutflow_yields}
Expected event yields at $3000~\mathrm{fb}^{-1}$ obtained by applying the
selection efficiencies of \cref{tab:cutflow_eff} to the signal and
background cross sections. The first row gives the normalised event yield
before the Delphes six-muon requirement. The
$N_\mu^{\rm reco}\geq6$ row includes detector reconstruction, acceptance,
and final muon isolation effects. For zero selected Monte Carlo events, the
quoted upper value is obtained by replacing zero by one event in the
corresponding sample.
Consistent with the efficiency definition, the initial signal yields
correspond to events containing at least six visible muons at truth level.}
\end{table}

The present Category-1 definition assumes equal dimuon masses on the two
branches and is therefore optimised for diagrams (B) and (D) of
\cref{fig:6mu-diags}. In diagram (C), one branch contains a $V'$ resonance
and the other an $H_D$ resonance. When their masses are well separated, the
equal-dimuon-mass condition reduces the reconstruction efficiency. The
limits obtained in this region are therefore conservative. A third category
allowing two different intermediate masses could recover these events, but
is not included in the present analysis.

%%%%%%%%%%%%%%%%%%%%%%%%%%%%%%%%%%%%%%%%%%%%%%%%%%%%%%%%%%%%%%%%%%%%%%%%%%%%%%%
\subsection{Constraints from existing Run-2 searches}
\label{subsec:current_limits}
%%%%%%%%%%%%%%%%%%%%%%%%%%%%%%%%%%%%%%%%%%%%%%%%%%%%%%%%%%%%%%%%%%%%%%%%%%%%%%%

We first determine the constraints from published LHC analyses in the
$(m_{V'},m_{H_D})$ plane using \texttt{CheckMATE}. The scan covers
\begin{equation}
m_{\mu'}=700,\ 750,\ 800,\ 850,\ 900~\GeV,
\end{equation}
and mediator masses between approximately $0.3$ and $316~\GeV$. For each
point, the full signal sample is passed through the corrected CMS detector
card and all implemented analyses.

This recast should be interpreted as a representative estimate of the current
Run-2 sensitivity rather than as a complete survey of all available LHC
multilepton searches. \texttt{CheckMATE} does not contain every recent ATLAS
and CMS analysis, and the aim here is not to reproduce the full experimental
programme. Instead, we use the analyses available in \texttt{CheckMATE} to
identify which existing search topology is most relevant for the MPVDM
six-muon signal and to quantify the approximate sensitivity of a
non-dedicated LHC search.

The strongest constraint throughout the scan is provided by the CMS
multilepton search CMS-SUS-16-039~\cite{CMS:2017moi}, based on
$35.9~\mathrm{fb}^{-1}$ at $\sqrt{s}=13~\TeV$. This search targets
electroweak production of charginos and neutralinos in final states with
two same-sign light leptons or at least three leptons, including categories
with electrons, muons, and hadronically decaying taus. It is relevant for the
MPVDM signal because the analysis contains multilepton signal regions with
very small SM backgrounds, and the six-muon signal can satisfy their lepton
requirements even though its underlying topology is not supersymmetric.

Among the \texttt{CheckMATE} signal regions, SR\_A44 is most frequently
selected as the most sensitive one in our scan. This region requires exactly
three light leptons, no hadronically decaying tau leptons, exactly one
opposite-sign same-flavour pair with invariant mass
$75<m_{\ell\ell}<105~\GeV$, no $b$-tagged jets, $\met>200~\GeV$, and a
transverse mass $M_T>160~\GeV$ formed from the third lepton and $\met$. It was
designed for high-mass chargino--neutralino production with leptonic $WZ$
decays. The large $\met$ and $M_T$ requirements suppress the SM $WZ$
background and make SR\_A44 a low-background, highly signal-like region.

Its sensitivity to the MPVDM signal is therefore understandable: six-muon
events can enter SR\_A44 when exactly three muons satisfy the CMS object and
signal-region requirements, while invisible $V_D$ states can provide large
$\met$. However, SR\_A44 neither requires six reconstructed muons nor uses
the repeated dimuon and trimuon resonance structure of the MPVDM cascade.

\begin{figure}[htbp]
\centering
\includegraphics[width=0.55\textwidth]{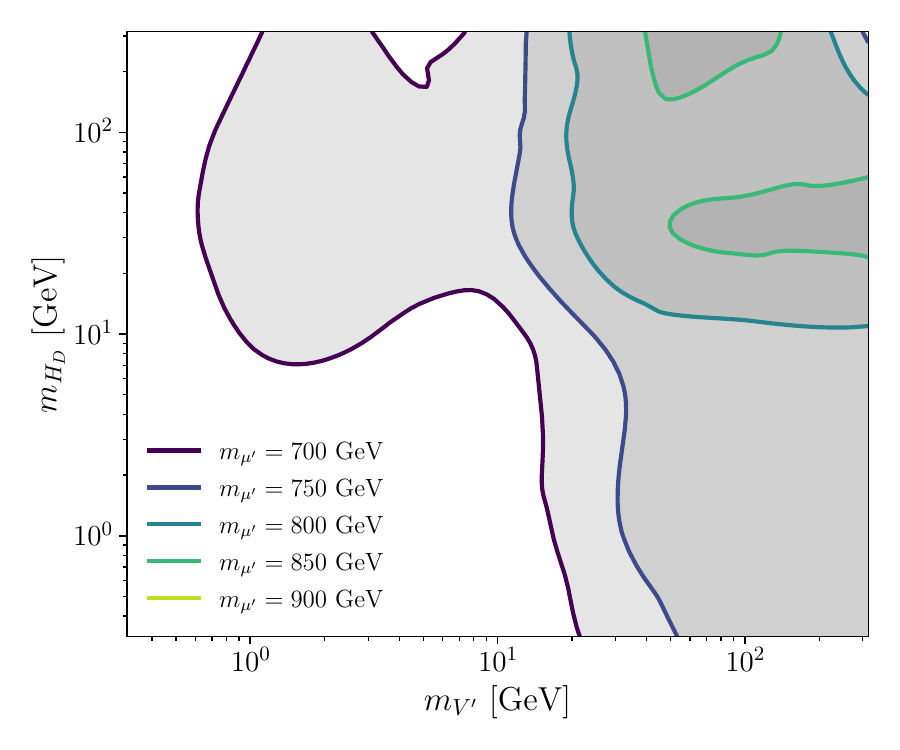}
\caption{\label{fig:cm_bounds}
Observed 95\% CL exclusion contours obtained with \texttt{CheckMATE} from
CMS-SUS-16-039 at $35.9~\mathrm{fb}^{-1}$. The contours $r=1$ are shown in
the $(m_{V'},m_{H_D})$ plane for the indicated values of $m_{\mu'}$.
For each mass, the darker side of the corresponding contour is
excluded and the lighter side is not excluded. For
$m_{\mu'}=900~\GeV$, the whole plane is non-excluded, so no yellow contour is visible.}
\end{figure}

The \texttt{CheckMATE} exclusion measure is
\begin{equation}
r=\frac{S-1.64\,\Delta S}{S_{95}^{\rm obs}},
\label{eq:checkmate_r}
\end{equation}
where $S$ and $\Delta S$ are the predicted signal yield and its Monte Carlo
uncertainty, and $S_{95}^{\rm obs}$ is the observed 95\% confidence-level
upper limit in the selected signal region. Points with $r>1$ are excluded.
The resulting contours are shown in \cref{fig:cm_bounds}. Existing generic
multilepton searches already test part of the low-$m_{\mu'}$ parameter
space, but their sensitivity varies strongly across the mediator-mass plane
because they were not designed for collimated dimuon resonances or repeated
trimuon parents.

For each value of $m_{\mu'}$, the corresponding contour separates a
darker excluded region from a lighter non-excluded region. If no contour is
visible for a mass shown in the legend, the whole displayed plane lies on
one side of the exclusion boundary. In particular, for
$m_{\mu'}=900~\GeV$ the entire plane is non-excluded, and is therefore shown
with the corresponding light shading without a visible yellow contour.

The selected CMS signal region has a very small effective background. This is
also visible from the \texttt{CheckMATE} response: parameter points with a
parton-level six-muon cross section of only
$(3$--$4)\times10^{-2}~\mathrm{fb}$ give $r\simeq0.24$, corresponding to
only $1.1$--$1.4$ parton-level events at $35.9~\mathrm{fb}^{-1}$. A signal
rate larger by a factor of about four would therefore approach the exclusion
boundary. This behaviour is consistent with a few-event Poisson limit rather
than a background-dominated Gaussian limit.

We do not infer the HL-LHC reach by rescaling the generic Run-2 recast. The
selected signal regions contain very few expected background events, so their
sensitivity is controlled by Poisson counting rather than by a
background-dominated Gaussian scaling. We therefore determine the
high-luminosity reach directly with the dedicated topology-based analysis
below.

%%%%%%%%%%%%%%%%%%%%%%%%%%%%%%%%%%%%%%%%%%%%%%%%%%%%%%%%%%%%%%%%%%%%%%%%%%%%%%%
\subsection{Expected reach of the dedicated six-muon search}
\label{subsec:dedicated_reach}
%%%%%%%%%%%%%%%%%%%%%%%%%%%%%%%%%%%%%%%%%%%%%%%%%%%%%%%%%%%%%%%%%%%%%%%%%%%%%%%

We next evaluate the expected sensitivity of the topology-based six-muon
search. The physical background yields are obtained by applying the sample
normalisations described in \cref{subsec:backgrounds} to the reconstruction
efficiencies.

The resulting background at $3000~\mathrm{fb}^{-1}$ is below
$10^{-4}$ events for Category~1 and below $10^{-2}$ events for Category~2
for target masses above $1200~\GeV$, and decreases further for larger target
masses. The numerical background yields are given in
\cref{app:selection_details}. Both categories, particularly Category~1, can
therefore be treated as effectively background-free in this mass range.

For a background-free counting experiment with zero observed events, the
95\% CL Poisson upper limit is close to three signal events. We consequently
define the expected exclusion boundary by
\begin{equation}
N_{\rm sig}=3.
\label{eq:three_event_limit}
\end{equation}
This criterion is applied after detector simulation and the complete
topology-based selection. In this regime, the excluded cross section scales
approximately as $1/\mathcal L$.

At a luminosity of $35.9~\mathrm{fb}^{-1}$, Category~1 provides the dominant
sensitivity over most of the mediator-mass plane. Category~2 contributes in
the region where the asymmetric $(1+5)\mu$ topology is appreciable, but does
not produce an independent exclusion contour for all masses.

\begin{figure}[htbp]
\centering
\includegraphics[width=0.75\textwidth]{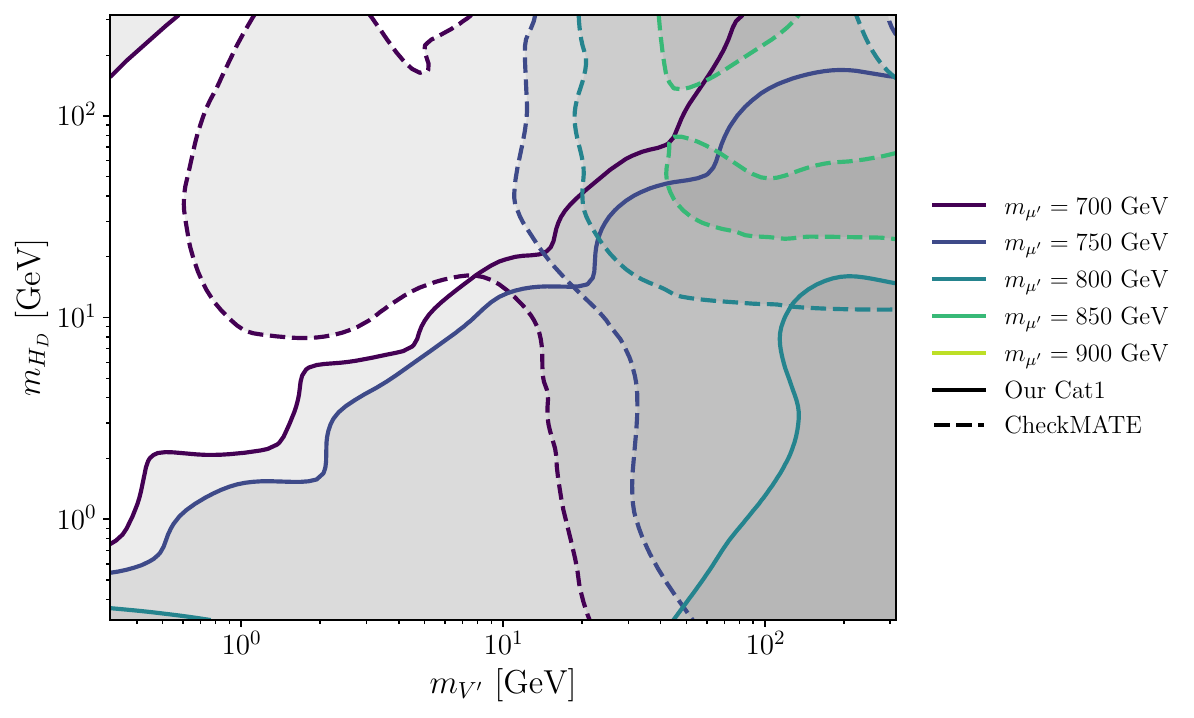}
\caption{\label{fig:fully_combined_limits}
Comparison between the expected dedicated Category~1 six-muon sensitivity
(solid curves) and the observed \texttt{CheckMATE} limits from
CMS-SUS-16-039 (dashed curves), both evaluated at
$35.9~\mathrm{fb}^{-1}$.
For each mass and analysis, the darker side of the corresponding
contour is excluded and the lighter side is not excluded. If a contour is absent for a given mass, it indicates that the entire displayed plane is not excluded.}
\end{figure}

The comparison with the observed \texttt{CheckMATE} constraint is shown in
\cref{fig:fully_combined_limits}. The dashed contours show the limits
obtained from CMS-SUS-16-039 through \texttt{CheckMATE}, while the solid
contours show the expected sensitivity of the dedicated Category~1
six-muon reconstruction at the same luminosity. CMS-SUS-16-039 remains
sensitive in regions where the six-muon branching fraction or the symmetric
Category~1 reconstruction efficiency is reduced. This is understandable
because SR\_A44 requires exactly three light leptons and therefore has a less
restrictive lepton-multiplicity requirement than the dedicated six-muon
selection.

The striking feature of \cref{fig:fully_combined_limits} is that the
dedicated six-muon analysis also excludes sizeable regions that are not
covered by the existing CMS multilepton search. This occurs, in particular,
in the lower-left part of the displayed plane and in the upper-left corner,
where the six-muon branching fraction and the Category~1 reconstruction
efficiency are favourable. Thus, even though the dedicated analysis imposes
a much stronger lepton-multiplicity requirement, it recovers parameter space
missed by the three-lepton CMS region because it exploits the repeated
resonance structure of the signal. Conversely, the CMS search retains
sensitivity where the six-muon branching fraction or Category~1 efficiency
is reduced. The two analyses therefore probe genuinely complementary regions
of parameter space, and a dedicated six-muon search would cover an important
gap left by the existing multilepton analysis.

\begin{figure}[htbp]
\centering
\begin{subfigure}{0.49\textwidth}
\includegraphics[width=\textwidth]{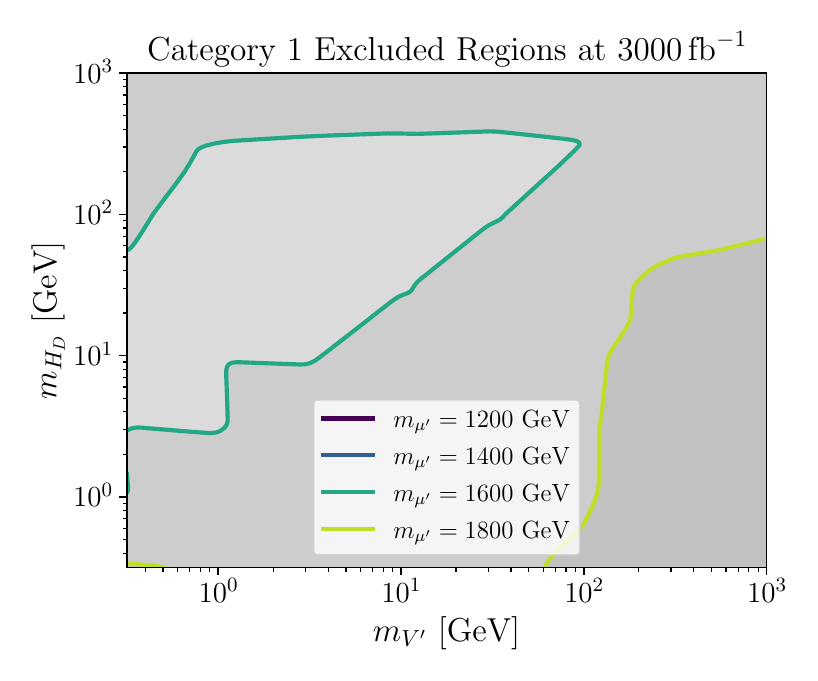}
\end{subfigure}
\hfill
\begin{subfigure}{0.49\textwidth}
\includegraphics[width=\textwidth]{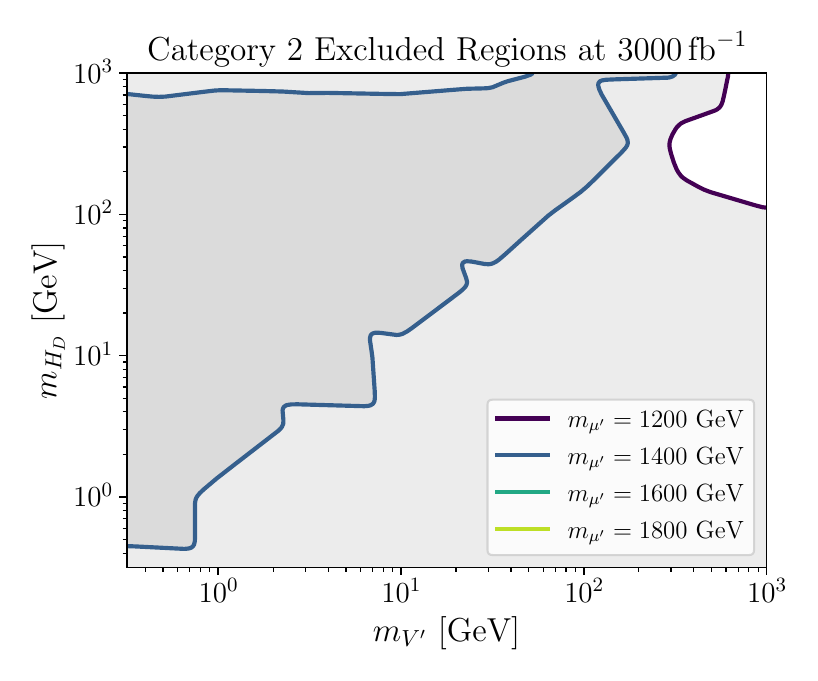}
\end{subfigure}
\caption{\label{fig:high_mass_combined_contours}
Expected exclusion contours for the high-mass region at
$3000~\mathrm{fb}^{-1}$, defined by $N_{\rm sig}>3$. The left panel shows
Category~1 and the right panel shows Category~2.
For each mass, the darker side of the corresponding contour is
excluded and the lighter side is not excluded. If no contour is visible, the
entire displayed plane lies on the same side of the exclusion boundary, as
indicated by the shading. In Category~1, the absence of the
$m_{\mu'}=1200$ and $1400~\GeV$ contours means that the whole plane is
excluded for these masses.
For $m_{\mu'}=1600$ and $1800~\GeV$, the exclusions remain only in the
high-yield regions. Category~2 gives a weaker but complementary reach where
the asymmetric $(1+5)\mu$ topology is enhanced.}
\end{figure}

At the HL-LHC, the expected reach is summarised in
\cref{fig:high_mass_combined_contours}. The contours correspond to
$N_{\rm sig}=3$ after the full detector-level and topology-based selection.
For each value of $m_{\mu'}$, the excluded region lies on the side of the
contour with the larger signal yield. In Category~1, the
$m_{\mu'}=1200$ and $1400~\GeV$ contours lie outside the displayed plane,
meaning that these masses are excluded throughout the scanned
$(m_{V'},m_{H_D})$ region. For $m_{\mu'}=1600$ and $1800~\GeV$, the reach
survives only in the high-yield parts of the plane, where both the six-muon
branching fraction and the Category~1 reconstruction efficiency are
favourable.

The Category~2 reach is weaker because the asymmetric $(1+5)\mu$ topology
has a smaller branching fraction over most of the plane and because only one
heavy parent is fully reconstructed. Nevertheless, it provides a useful
cross-check and a complementary probe of the region where the $(1+5)\mu$
contribution is enhanced. We therefore do not combine Category~2 with
Category~1 when defining the main mass reach, but retain it as an independent
handle on the underlying event topology.

\begin{figure}[htbp]
\centering
\begin{subfigure}{0.65\textwidth}
\includegraphics[width=\textwidth]{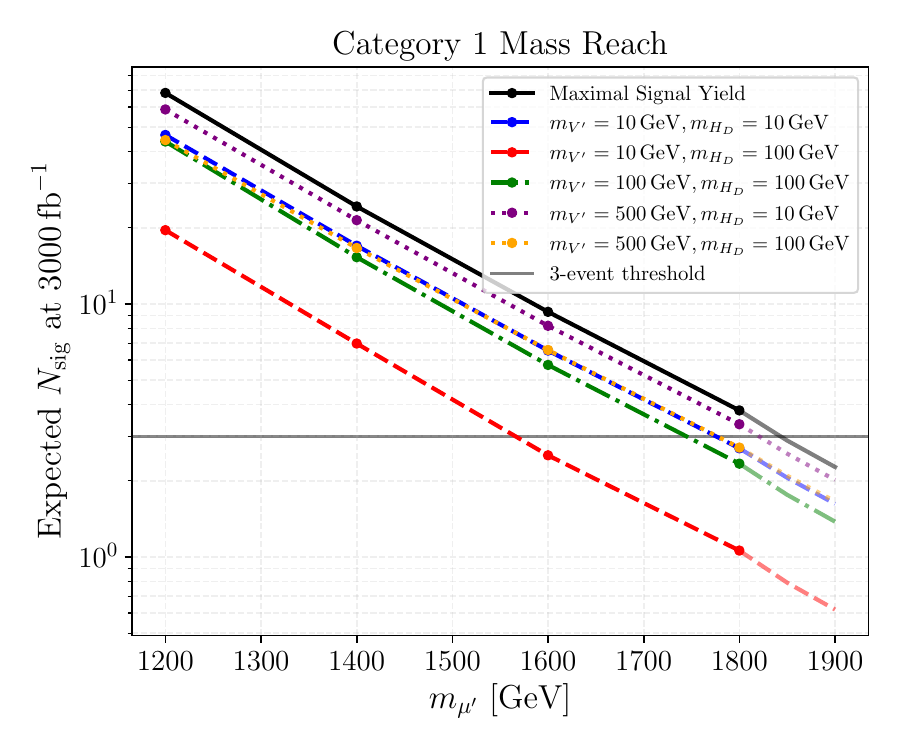}
\label{fig:hl_cat1_massreach}
\end{subfigure}
\\
\begin{subfigure}{0.65\textwidth}
\includegraphics[width=\textwidth]{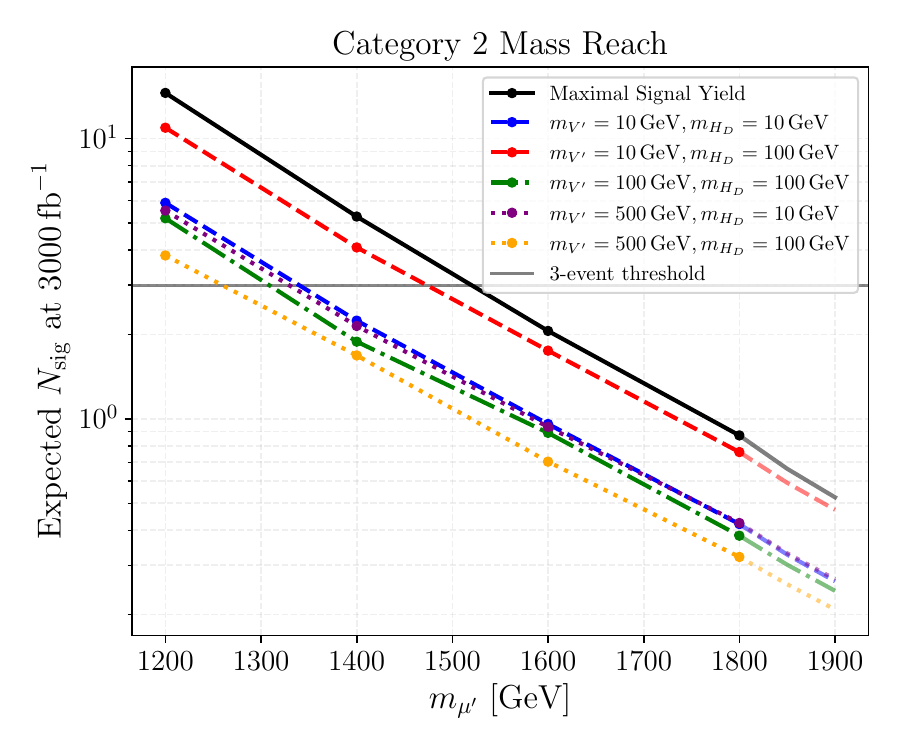}
\label{fig:hl_cat2_massreach}
\end{subfigure}
\caption{\label{fig:hl_massreach}
Expected signal yield at $3000~\mathrm{fb}^{-1}$ as a function of
$m_{\mu'}$ for representative mediator masses. The top panel shows the
sensitivity of Category~1, while the bottom panel shows the sensitivity of
Category~2. The horizontal line marks the three-event exclusion threshold,
and the ``maximal signal'' curve gives the largest yield found in each
$m_{\mu'}$ slice. The displayed range starts at
$m_{\mu'}=1.2~\TeV$, matching the high-mass HL-LHC scan.}
\end{figure}

The mass dependence is displayed more directly in
\cref{fig:hl_massreach} for representative mediator masses. Category~1
retains a three-event reach up to approximately $1.9~\TeV$ for favourable
spectra. Category~2 falls more rapidly with $m_{\mu'}$ and acts mainly as a
complementary topology channel rather than setting the overall mass reach.

The detailed detector and reconstruction efficiencies underlying these
projections are provided in \cref{app:selection_details}. They show that the
parton-to-detector six-muon efficiency is high and only weakly dependent on
$m_{\mu'}$, while the topology-dependent efficiency is controlled mainly by
the relation between $m_{V'}$ and $m_{H_D}$. This confirms that the eventual
mass reach is determined primarily by the rapidly falling Drell--Yan
production cross section rather than by detector acceptance.
%%%%%%%%%%%%%%%%%%%%%%%%%%%%%%%%

%%%%%%%%%%%%%%%%%%%%%%%%%%%%%
%%%%%%%%%%%%%%%%%%%%%%%%%%%%%

\section{Conclusion}
\label{sec:conclusion}

We have performed a collider study of a novel six-muon signature arising
from vector-like lepton pair production followed by cascade decays through a
dark sector. In the muonic fermionic portal to vector dark matter, the signal
originates from Drell--Yan production,
$pp\to\mu'^{+}\mu'^{-}$, followed by decays through the dark vector $V'$
and the dark scalar $H_D$. The same six-muon multiplicity can be produced by
both the symmetric $(3+3)\mu$ topology and the asymmetric $(1+5)\mu$
topology, and both contributions have been included in the analysis.

The six-muon channel is the most powerful multimuon target in this model. It
remains sizeable over a broader region of parameter space than the eight- and
ten-muon channels, while being much cleaner and more reconstructable than the
four-muon channel. More generally, it illustrates a class of six-lepton, or
six-fermion, collider topologies that can arise from pair-produced
vector-like fermions. Such final states are not merely spectacular counting
signatures: their correlated resonance structures can reveal the underlying
pattern of new particles and interactions.

We developed an explicit topology-based reconstruction strategy for this
signature using two exclusive categories. Category~1 targets the symmetric
$(3+3)\mu$ topology by partitioning the event into two compatible trimuon
systems. Category~2 targets the asymmetric $(1+5)\mu$ topology by
reconstructing one visible five-muon branch while allowing the other branch
to contain invisible dark-sector particles. The repeated dimuon, trimuon,
and five-muon mass structures are used to resolve the combinatorics and
reconstruct the intermediate $V'$ or $H_D$ states and the parent vector-like
muon. The signal is therefore treated as a reconstructable cascade topology
rather than as a generic multilepton excess.

We simulated the signal and Standard Model backgrounds at detector level,
including a dedicated treatment of rare heavy-flavour muons. After the
six-muon requirement and topology-based reconstruction, the expected
background is negligible at the HL-LHC. The sensitivity is consequently
controlled primarily by the signal rate and reconstruction efficiency rather
than by background rejection.

Existing Run-2 multilepton searches already constrain part of the low-mass
parameter space, but they do not exploit the repeated resonance structure of
the signal. A dedicated six-muon analysis therefore provides a substantial
gain in sensitivity and probes regions not covered by the existing inclusive
multilepton search. At the HL-LHC, the reach extends up to approximately
$1.9~\TeV$ for favourable regions of the $(m_{V'},m_{H_D})$ plane.

An important feature of the dominant symmetric $(3+3)\mu$ topology is that
it is fully visible and does not require  missing transverse momentum.
This differs from many supersymmetry-motivated multilepton searches, which
rely strongly on $\met$ to suppress Standard Model backgrounds. Applying
such requirements to the present signal would remove precisely the region in
which both vector-like-lepton decay chains can be reconstructed. Searches for
high-multiplicity leptonic final states should therefore remain inclusive in
$\met$ and instead exploit their resonance and cascade structure. The
asymmetric $(1+5)\mu$ topology can contain genuine missing momentum, but it
should be treated as a complementary category rather than used to define the
whole search.

The implications of this study extend beyond the specific MPVDM model.
Pair-produced vector-like fermions can naturally generate six-fermion
cascade topologies, and these final states can carry sufficient internal
structure to reconstruct the new sector. Dedicated searches for such
topology-rich multilepton events would provide a clean and powerful probe of
fermionic portals to dark gauge sectors and of other BSM scenarios that may
be weakly constrained by standard inclusive searches. The negligible
background and explicit reconstruction strategy presented here make the
six-muon signature a well-motivated target for future ATLAS and CMS
multilepton analyses.
%%%%%%%%%%%%%%%%%%%%%%%%%%%%%%%%%%

\section*{Acknowledgement}

AB and CY are supported in part through the NExT Institute and STFC CG ST/X000583/1.
AB acknowledges support from the Leverhulme Trust project MONDMag (RPG-2022-57).
MC thanks the University of Southampton for hosting her during this work.
We acknowledge the use of the IRIDIS High Performance Computing Facility and associated support services at the University of Southampton in the completion of this work.

%%%%%%%%%%%%%%%%%%%%%%%%%%%%%%%%
%%%%%%%%%%%%%%%%%%%%%%%%%%%%%%%%

%%%%%%%%%%%%%%%%%%%%%%%%%%%%%%%%%%%%%%%%%%%%%%%%%%%%%%%%%%%%%%%%%%%%%%%%%%%%%%%
\clearpage
\appendix
%%%%%%%%%%%%%%%%%%%%%%%%%%%%%%%%%%%%%%%%%%%%%%%%%%%%%%%%%%%%%%%%%%%%%%%%%%%%%%%

\section{Heavy-flavour background generation and normalisation}

\label{app:background_normalisation}
%%%%%%%%%%%%%%%%%%%%%%%%%%%%%%%%%%%%%%%%%%%%%%%%%%%%%%%%%%%%%%%%%%%%%%%%%%%%%%%

The $t\bar t$, $t\bar t Z$, and four-top backgrounds require additional
muons from heavy-flavour decays to enter the six-muon signal region. Since
the probability of obtaining the required muon multiplicity is small, direct
generation with unrestricted heavy-flavour decays would lead to poor Monte
Carlo statistics.

We therefore developed dedicated code interfacing \pythia{} with
\texttt{Delphes}. For each showered and hadronised hard event, the
weakly decaying bottom hadrons are initially kept stable. The resulting event
configuration is stored, after which the bottom hadrons are decayed repeatedly
using the \pythia{} decay tables. The subsequent decays of their unstable
descendants, including charm hadrons, are included in each decay trial.

The repeated-decay code can call either \texttt{moreDecays(false)} or
\texttt{moreDecays(true)}. The production samples use
\texttt{moreDecays(false)}, which is the default treatment in the \pythia{}
version used here. In this mode, decay systems containing partonic daughters
are converted into hadronic final states within the decay procedure rather
than retained as unresolved partons before being passed to \texttt{Delphes}.
We verified that \texttt{moreDecays(true)} gives filtering efficiencies
consistent with those obtained using the older \pythia~8.2 implementation.
All results quoted in this work use \texttt{moreDecays(false)}.

After each decay trial, final-state muons descending from the original
bottom hadrons are counted. This includes direct muons from bottom-hadron
decays and secondary muons produced in subsequent decay chains. The required
additional-muon multiplicities are
\begin{equation}
N_{\mu}^{\rm HF}\geq
\begin{cases}
4, & t\bar t,\\
2, & t\bar t Z,\\
2, & t\bar t t\bar t.
\end{cases}
\label{eq:hf_muon_requirements}
\end{equation}
For each hard event, the heavy-flavour decays are repeated for a fixed
maximum number of trials. Every decay configuration satisfying the required
additional-muon multiplicity is passed to \texttt{Delphes}, until the target
number of written events is reached.

Let $N_{\rm trial}$ denote the total number of generated heavy-flavour decay
configurations and $N_{\rm acc}$ the number satisfying the required
additional-muon multiplicity. The filtering efficiency is
\begin{equation}
\epsilon_{\rm HF}=\frac{N_{\rm acc}}{N_{\rm trial}}.
\label{eq:hf_efficiency_appendix}
\end{equation}

For a sample generated in several statistically independent parts, the
combined efficiency is calculated as
\begin{equation}
\epsilon_{\rm HF}^{\rm combined}=\frac{\sum_i N_{{\rm acc},i}}{\sum_i N_{{\rm trial},i}}.
\label{eq:hf_efficiency_combined_appendix}
\end{equation}

The cross section represented by the filtered sample is
\begin{equation}
\sigma_{\rm filter}=\sigma_{\rm ME}\frac{N_{\rm acc}}{N_{\rm trial}}=\sigma_{\rm ME}\epsilon_{\rm HF},
\label{eq:hf_filtered_xsec_appendix}
\end{equation}
where $\sigma_{\rm ME}$ is the cross section of the explicitly generated
matrix-element final state before heavy-flavour decays.

All accepted events carry the same statistical weight. If $N_{\rm sel}$
events survive the detector-level selection, their corresponding physical
cross section is
\begin{equation}
\sigma_{\rm sel}=\sigma_{\rm ME}\frac{N_{\rm sel}}{N_{\rm trial}}.
\label{eq:hf_selected_xsec_appendix}
\end{equation}
The normalisation is therefore determined by the total number of decay
trials and not by assigning a separate inverse trial count to each hard
event.

This procedure is more complete than forcing selected $B$- and $D$-hadron
decays to contain muons and correcting the resulting sample using approximate
semileptonic branching fractions. For example,
\begin{equation}
B_{bc}=\Br(b\to\mu\nu_\mu X_c)\Br(c\to\mu\nu_\mu X_s),
\end{equation}
with
\begin{equation}
\Br(b\to\mu\nu_\mu X_c)\simeq0.105,\qquad
\Br(c\to\mu\nu_\mu X_s)\simeq0.090,
\end{equation}
gives
\begin{equation}
B_{bc}\simeq9.45\times10^{-3}.
\end{equation}
The estimate $B_{bc}^2$ accounts only for the restricted topology in which
each of the two primary bottom-hadron decay chains produces one direct muon
and one secondary muon from a charm-hadron decay. It does not include all
decay chains contributing to the required additional-muon multiplicity.

The repeated-decay simulation includes direct semileptonic bottom-hadron
decays, secondary muons from charm-hadron decays, bottom-hadron decays
containing more than one charm hadron, muons from charmonium and tau decays,
different weakly decaying bottom-hadron species, and additional heavy flavour
generated in the parton shower. It therefore evaluates directly the
inclusive probability relevant to the six-muon analysis.

For the samples used in this work, the resulting filtering efficiencies are
\begin{align}
\epsilon_{\rm HF}(t\bar t) &= 3.0\times10^{-11}, \\
\epsilon_{\rm HF}(t\bar t Z) &= 5.49\times10^{-5}, \\
\epsilon_{\rm HF}(t\bar t t\bar t) &= 2.38\times10^{-4}.
\end{align}
The corresponding filtered cross sections are
\begin{align}
\sigma_{\rm filter}(t\bar t)
&= 2.736\times10^{3}~{\rm fb}\times3.0\times10^{-11}
= 8.21\times10^{-8}~{\rm fb}, \\
\sigma_{\rm filter}(t\bar t Z)
&= 0.248~{\rm fb}\times5.49\times10^{-5}
= 1.36\times10^{-5}~{\rm fb}, \\
\sigma_{\rm filter}(t\bar t t\bar t)
&= 1.41\times10^{-3}~{\rm fb}\times2.38\times10^{-4}
= 3.36\times10^{-7}~{\rm fb}.
\end{align}

The dedicated code implementing the repeated heavy-flavour decays and the
additional-muon filter is provided as supplementary material, together with
the source file and Makefile required for compilation within
\texttt{Delphes}.

%%%%%%%%%%%%%%%%%%%%%%%%%%%%%%%%%%%%%%%%%%%%%%%%%%%%%%%%%%%%%%%%%%%%%%%%%%%%%%%
\section{Invariant-mass uncertainties and reconstruction variables}

\label{app:chi2_details}
%%%%%%%%%%%%%%%%%%%%%%%%%%%%%%%%%%%%%%%%%%%%%%%%%%%%%%%%%%%%%%%%%%%%%%%%%%%%%%%

For a system of $n$ reconstructed particles with total four-momentum
$P^\mu_{\rm tot}=\sum_kp_k^\mu$, we write
$M^2=E_{\rm tot}^2-|\vec p_{\rm tot}|^2$. Assuming an uncertainty
$\sigma_k=fE_k$ for each four-momentum component and neglecting correlations,
error propagation gives
\begin{equation}
\sigma_M^2
=
\frac{f^2(E_{\rm tot}^2+|\vec p_{\rm tot}|^2)}{M^2}
\sum_{k=1}^{n}E_k^2 ,
\label{eq:mass_variance_appendix}
\end{equation}
and therefore
\begin{equation}
\left(\frac{\sigma_M}{M}\right)^2
=
\frac{f^2(E_{\rm tot}^2+|\vec p_{\rm tot}|^2)}{M^4}
\sum_{k=1}^{n}E_k^2 .
\label{eq:relative_mass_variance_appendix}
\end{equation}

For two independently reconstructed masses, the compatibility variable is
\begin{equation}
\chi^2(M_1,M_2)
=
\frac{\left[\ln(M_1)-\ln(M_2)\right]^2}
{\left(\sigma_{M_1}/M_1\right)^2+
 \left(\sigma_{M_2}/M_2\right)^2}.
\label{eq:chi2_two_masses_appendix}
\end{equation}
For a reconstructed mass $M$ compared with a target resonance
$M_0\pm\sigma_{M_0}$, we use
\begin{equation}
\chi^2(M,M_0)
=
\frac{\left[\ln(M)-\ln(M_0)\right]^2}
{\left(\sigma_M/M\right)^2+
 \left(\sigma_{M_0}/M_0\right)^2}.
\label{eq:chi2_target_appendix}
\end{equation}
The quantities in \cref{eq:chi2_two_masses_appendix,eq:chi2_target_appendix} are compatibility variables rather than fitted likelihoods. A small value means that the two candidate masses are mutually compatible, or that a reconstructed mass is compatible with the target value. The logarithm is used because the relevant question is fractional mass agreement: a fixed absolute mass difference should not be weighted in the same way for a GeV-scale dimuon resonance and for a TeV-scale parent. The terms entering \cref{eq:chi2_cat1,eq:chi2_cat2} are then added to form the total assignment score.

%%%%%%%%%%%%%%%%%%%%%%%%%%%%%%%%%%%%%%%%%%%%%%%%%%%%%%%%%%%%%%%%%%%%%%%%%%%%%%%
\section{Selection efficiencies and background yields}

\label{app:selection_details}
%%%%%%%%%%%%%%%%%%%%%%%%%%%%%%%%%%%%%%%%%%%%%%%%%%%%%%%%%%%%%%%%%%%%%%%%%%%%%%%

\begin{figure}[htbp]
\centering
\includegraphics[width=0.9\linewidth]{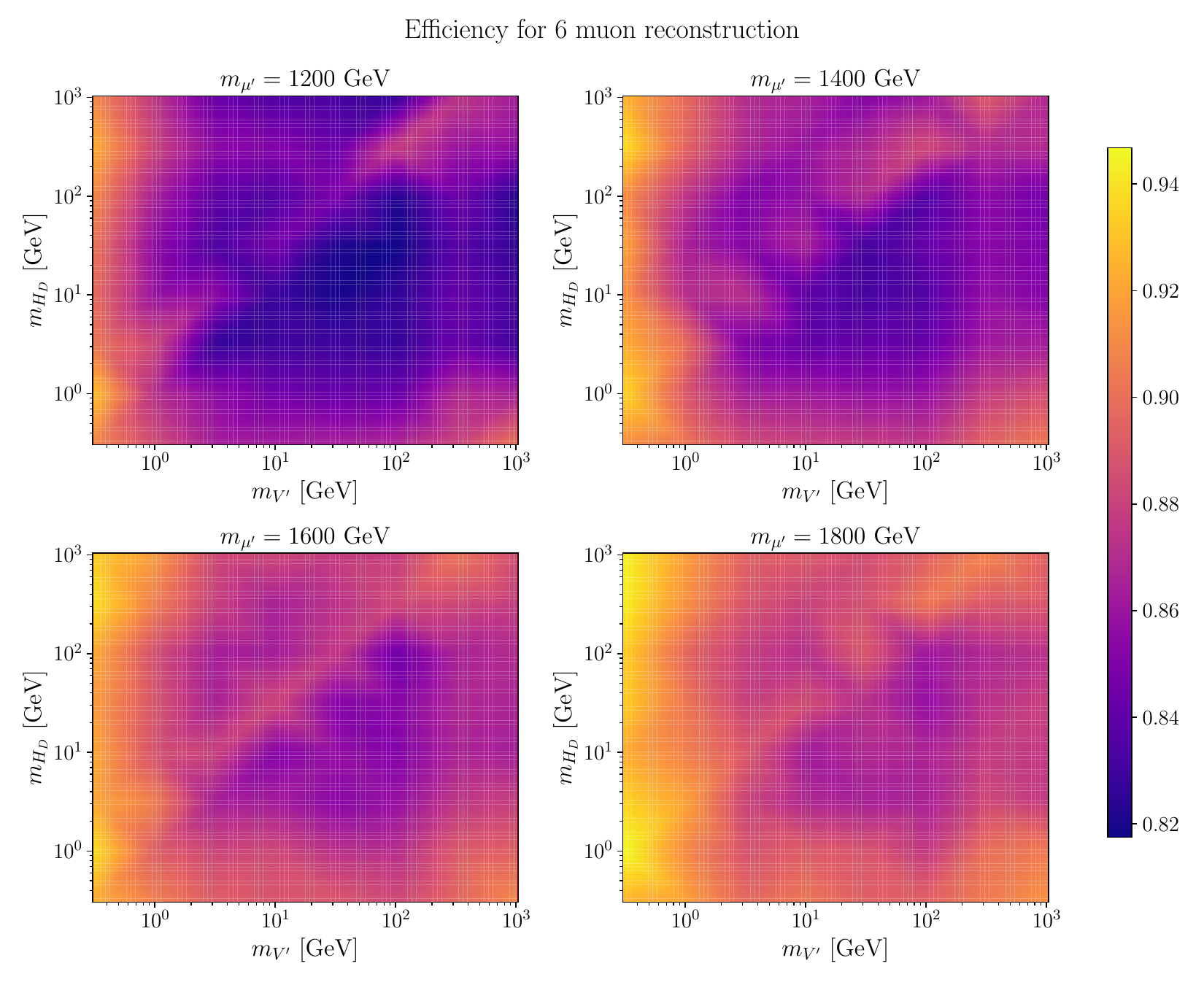}
\caption{Efficiency for 6 muon reconstruction in the $m_{V'}$--$m_{H_D}$ parameter space. The four panels correspond to the target heavy-lepton masses $m_{\mu'} =$ 1200, 1400, 1600, and 1800~GeV. The colormap indicates the fraction of generated events with at least six reconstructed muons.}
\label{fig:eff_filter}
\end{figure}

\begin{figure}[htbp]
\centering
\includegraphics[width=0.9\linewidth]{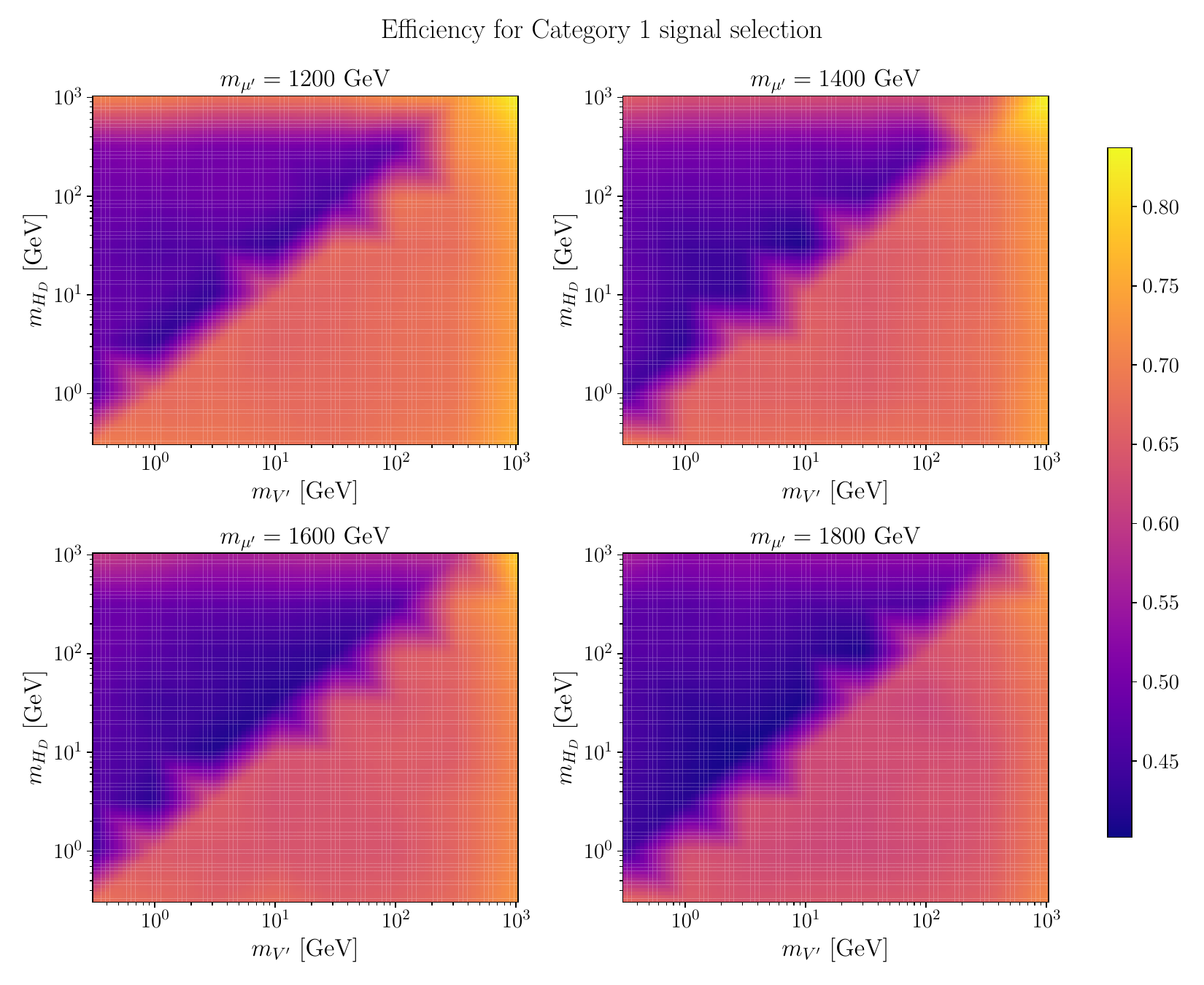}
\caption{Efficiency for Category 1 signal selection in the $m_{V'}$--$m_{H_D}$ parameter space. The four panels correspond to the target heavy-lepton masses $m_{\mu'} =$ 1200, 1400, 1600, and 1800~GeV. The efficiency is defined relative to the number of events passing the six-muon filter.}
\label{fig:eff_cat1}
\end{figure}

\begin{figure}[htbp]
\centering
\includegraphics[width=0.9\linewidth]{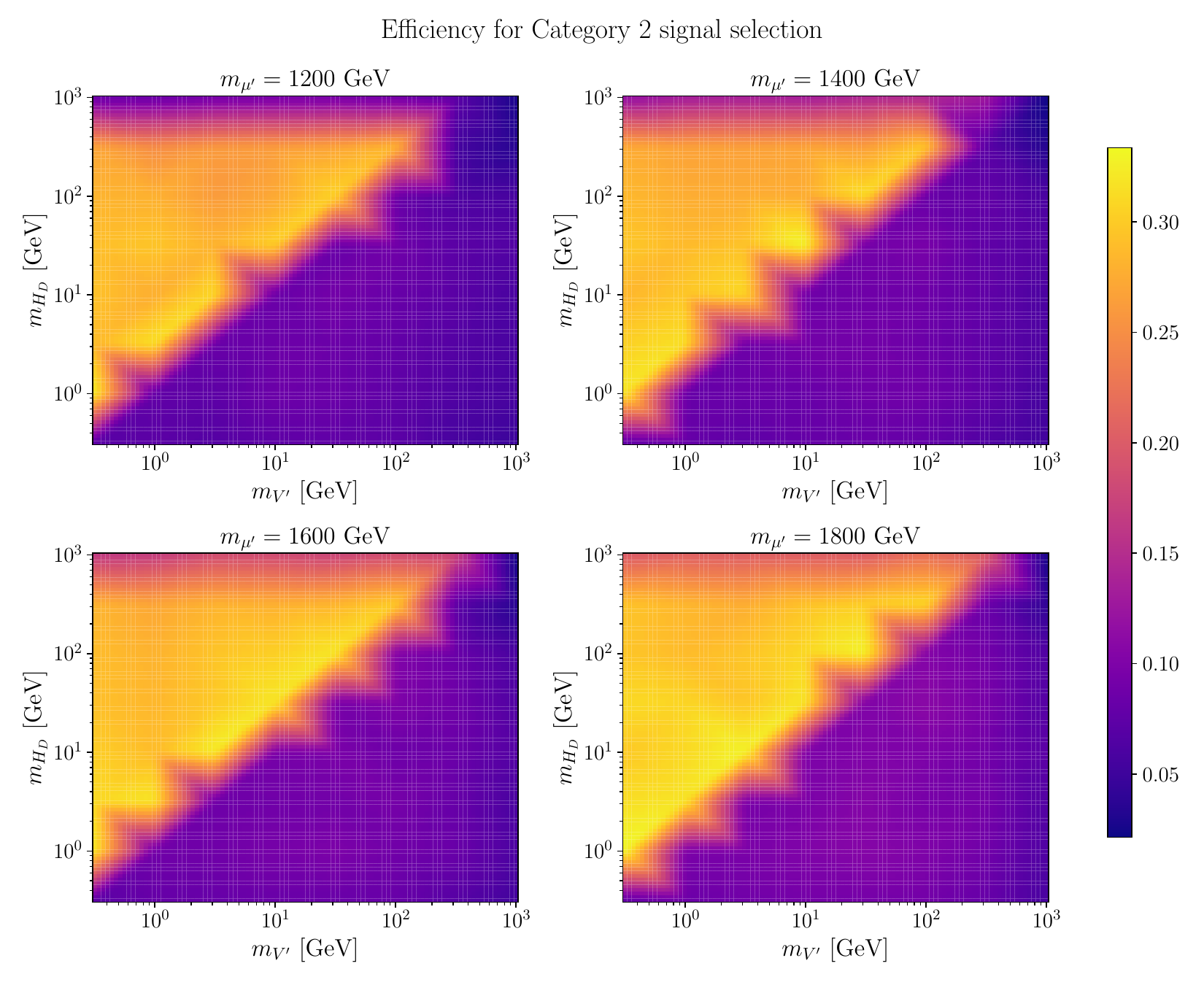}
\caption{Efficiency for Category 2 signal selection in the $m_{V'}$--$m_{H_D}$ parameter space. The four panels correspond to the target heavy-lepton masses $m_{\mu'} =$ 1200, 1400, 1600, and 1800~GeV. The efficiency is defined relative to the number of events passing the six-muon filter.}
\label{fig:eff_cat2}
\end{figure}

We define the parton-to-detector efficiency as the number of events with at
least six reconstructed muons divided by the number containing at least six
muons in the LHE record. The topology efficiencies are defined relative to
the same parton-level denominator. The maps in
\cref{fig:eff_filter,fig:eff_cat1,fig:eff_cat2} show these efficiencies in
the $(m_{V'},m_{H_D})$ plane.

The detector-level six-muon efficiency remains near $80$--$90\%$ over the
displayed mass range, as shown in \cref{fig:eff_filter}. The Category-1 and
Category-2 efficiencies in \cref{fig:eff_cat1,fig:eff_cat2} are lower because
they include the topology assignment and the resonance requirements. The
corresponding background yields after the successive selections are summarised
in \cref{tab:bkg_yields}.

\begin{table}[htbp]
\centering
\setlength{\tabcolsep}{5.0pt}
\renewcommand{\arraystretch}{1.15}
\begin{tabular}{lccccc}
\toprule
Selection stage
& $6\mu$
& $ZZWW$
& $t\bar t Z$
& $4t$
& Total \\
\midrule
Initial
& $2.59\times10^{-1}$
& $1.61\times10^{-2}$
& $4.09\times10^{-2}$
& $1.01\times10^{-3}$
& $3.17\times10^{-1}$ \\
$N_\mu\geq6$
& $2.04\times10^{-1}$
& $8.66\times10^{-3}$
& $2.24\times10^{-5}$
& $3.04\times10^{-7}$
& $2.13\times10^{-1}$ \\
\midrule
Cat.~1, $m_{\mu'}=1200~\GeV$
& $2.22\times10^{-5}$ & $1.62\times10^{-5}$ & $<1.49\times10^{-7}$ & $<1.01\times10^{-9}$ & $3.84\times10^{-5}$ \\
Cat.~1, $m_{\mu'}=1400~\GeV$
& $3.70\times10^{-6}$ & $6.92\times10^{-6}$ & $<1.49\times10^{-7}$ & $<1.01\times10^{-9}$ & $1.06\times10^{-5}$ \\
Cat.~1, $m_{\mu'}=1600~\GeV$
& $<3.70\times10^{-6}$ & $3.22\times10^{-6}$ & $<1.49\times10^{-7}$ & $<1.01\times10^{-9}$ & $3.22\times10^{-6}$ \\
Cat.~1, $m_{\mu'}=1800~\GeV$
& $<3.70\times10^{-6}$ & $1.29\times10^{-6}$ & $<1.49\times10^{-7}$ & $<1.01\times10^{-9}$ & $1.29\times10^{-6}$ \\
\midrule
Cat.~2, $m_{\mu'}=1200~\GeV$
& $2.49\times10^{-2}$ & $2.13\times10^{-3}$ & $<1.49\times10^{-7}$ & $8.06\times10^{-9}$ & $2.70\times10^{-2}$ \\
Cat.~2, $m_{\mu'}=1400~\GeV$
& $1.29\times10^{-2}$ & $1.49\times10^{-3}$ & $<1.49\times10^{-7}$ & $3.02\times10^{-9}$ & $1.44\times10^{-2}$ \\
Cat.~2, $m_{\mu'}=1600~\GeV$
& $6.84\times10^{-3}$ & $1.04\times10^{-3}$ & $<1.49\times10^{-7}$ & $2.02\times10^{-9}$ & $7.88\times10^{-3}$ \\
Cat.~2, $m_{\mu'}=1800~\GeV$
& $3.78\times10^{-3}$ & $7.06\times10^{-4}$ & $<1.49\times10^{-7}$ & $1.01\times10^{-9}$ & $4.48\times10^{-3}$ \\
\bottomrule
\end{tabular}
\caption{\label{tab:bkg_yields}
Expected background yields at $3000~\mathrm{fb}^{-1}$. The first two rows
show the yields before topology-dependent reconstruction. The remaining
rows show the yields after the Category~1 and Category~2 selections for
the target masses used in the HL-LHC analysis. The heavy-flavour
backgrounds include the corresponding additional-muon filtering
efficiencies.}
\end{table}

\clearpage

\bibliographystyle{apsrev4-1}
\bibliography{ref}

\end{document}